\documentclass[twocolumn,english,asp,prb,showpacs,preprintnumbers]{revtex4}
\usepackage[T1]{fontenc}
\usepackage[latin9]{inputenc}
\usepackage{color}
\usepackage{float}
\usepackage{amsmath}
\usepackage{graphicx}
\usepackage{amssymb}
\usepackage{esint}

\makeatletter
\@ifundefined{textcolor}{}
{%
 \definecolor{BLACK}{gray}{0}
 \definecolor{WHITE}{gray}{1}
 \definecolor{RED}{rgb}{1,0,0}
 \definecolor{GREEN}{rgb}{0,1,0}
 \definecolor{BLUE}{rgb}{0,0,1}
 \definecolor{CYAN}{cmyk}{1,0,0,0}
 \definecolor{MAGENTA}{cmyk}{0,1,0,0}
 \definecolor{YELLOW}{cmyk}{0,0,1,0}
 }

\@ifundefined{definecolor}
 {\@ifundefined{definecolor}
 {\@ifundefined{definecolor}
 {\usepackage{color}}{}
}{}
}{}

\makeatother
\usepackage{babel}

\begin{document}

\title{An equivalent expression of $Z_{2}$ Topological Invariant for band
insulators using Non-Abelian Berry's connection }

\author{Rui Yu$^{1}$, Xiao Liang Qi$^{2}$, Andrei Bernevig$^{3}$, Zhong
Fang$^{1}$ and Xi Dai$^{1}$}

\affiliation{$^{1}$Beijing National Laboratory for Condensed Matter Physics and
Institute of Physics, Chinese Academy of Sciences, Beijing 100080,
China}

\affiliation{$^{2}$Department of Physics, Stanford University, Stanford, California
94305, USA}

\affiliation{$^{3}$ Department of Physics, Princeton University, Princeton, New
Jersey 08540, USA}

\date{\today}
\begin{abstract}
We introduce a new expression for the $Z_{2}$ topological invariant
of band insulators using non-Abelian Berry's connection. Our expression
can identify the topological nature of a general band insulator \emph{without}
any of the gauge fixing problems that plague the concrete implementation
of previous invariants. The new expression can be derived from the
\textquotedbl{}partner switching\textquotedbl{} of the Wannier function
center during time reversal pumping and is thus equivalent to the
$Z_{2}$ topological invariant proposed by Kane and Mele.
\end{abstract}


\maketitle

\section{Introduction}

Topological invariants play a very important role in the classification
of band insulators. The studies on integer quantum Hall effect (IQHE)
show that 2D band insulators without time reversal symmetry can be
classified by the Chern number - an integer describing the topological
structure of a set of fully occupied Bloch bands without Kramers degeneracy.
Systems with non-zero Chern number exhibit IQHE\cite{laughlin_quantized_1981,thouless_quantized_1982}.

A similar idea can be also applied to band insulators with time reversal
symmetry. Recently a $Z_{2}$ topological invariant has been proposed
by Kane and Mele to characterize the time reversal invariant band
insulators in 2D\cite{kane_quantum_2005,kane_z_2_2005}. According
to this new topological invariant, all the 2D band insulators with
time-reversal invariance can be divided into two classes. The normal
insulators with even $Z_{2}$ number and topological insulators with
odd $Z_{2}$ number \cite{moore_topological_2007,roy_topological_2009,fu_time_2006}.
The 2D topological insulators will exhibit a quantum spin Hall effect
(QSHE)\cite{kane_quantum_2005,bernevig_quantum_2006}, which is characterized
by the presence of helical edge states \cite{kane_quantum_2005,wu_helical_2006,xu_stability_2006,roth_nonlocal_2009,
konig_quantum_2007,bernevig_quantum_2006,bernevig_quantum_2006-1,qi_topological_2006}.
Interestingly the $Z_{2}$ topological invariant can also be generalized
to the 3D band insulators with time reversal symmetry\cite{fu_topological_2007,roy_topological_2009,moore_topological_2007}.
In this case, there are four independent $Z_{2}$ topological numbers:
one strong topological index and three weak topological indices\cite{fu_topological_2007,fu_topological_2007-1,teo_surface_2008,
fukui_quantum_2007,ran2009}.
The 3D time reversal invariant band insulators can be classified as
normal insulators, weak topological insulators (WTI) and strong topological
insulators (STI) according to the values of these four $Z_{2}$ topological
indices. Among them, the STI attracts much attention due to its unique
Dirac type surface states and robustness against disorder \cite{zhang_topological_2009,chen_experimental_2009,hsieh_topological_2008,
xia_observation_2009,analytis_bulk_2010,hsieh_observation_2009-1,
park_quasiparticle_2010,shen_quantum_2009,yan_theoretical_2010,
lin_half-heusler_2010,hasan_topological_2010,shan_effective_2010}.
The helical spin structure of the Dirac type surface states has been
experimentally implied by the standing wave structure in STM images
around an impurity scattering center and measured directly by spin
resolved angle resolved photo emission spectra (ARPES)\cite{roushan_topological_2009,alpichshev_stm_2010,
lee_quasiparticle_2009,zhang_crossover_2010,zhang_experimental_2009,
zhou_theory_2009,guo_theory_2010,hsieh_tunable_2009,hsieh_observation_2009}.
The Dirac type 2D electron gas living on the surface of STI or at
the interface between STI and normal insulators provides a new playground
for spintronics and quantum computing.

Since the $Z_{2}$ invariant characterizes whether a system is topologically
trivial or nontrivial, its computation is essential to the field of
topological insulators. For band insulators with extra spacial inversion
symmetry, the $Z_{2}$ topological numbers can be easily computed
as the product of half of the parity (Kramers pairs have identical
parities) numbers for all the occupied states at the high symmetry
points\cite{fu_topological_2007-1}. The situation becomes complicated
in the general case where spacial inversion symmetry is absent. At
the present, numerically there are three different ways to judge if
a band insulator without inversion symmetry is a TI or not. i) Compute
the $Z_{2}$ numbers using the integration of both Berry's connection
and curvature over half of the Brillouin Zone (BZ). In order to do so, one has to set
up a mesh in the k-space and calculate the corresponding quantities
on the lattice version of the problem\cite{fukui_chern_2005,fukui_quantum_2007}.
Since the calculation involves the Berry's connection, one has to
numerically fix the gauge on the half BZ, which is not easy for the
realistic wave functions obtained by first principle calculation.
ii) Start from an artificial system with spacial inversion symmetry,
and then smoothly \textquotedbl{}deform\textquotedbl{} the Hamiltonian
towards the realistic one without inversion symmetry. If the energy
gap never closes at any points in the BZ during the \textquotedbl{}deformation\textquotedbl{}
process, the realistic system must share the same topological nature
with the initial reference system whose $Z_{2}$ number can be easily
counted by the inversion eigenvalue formula. Unfortunately making
sure that the energy gap remains open on the whole BZ is very difficult
numerically, especially in 3D. iii) Directly calculate the surface
states. For most of the TI materials, the first principle calculation
for the surface states is numerically heavy. Therefore it is valuable
to develop a mathematically equivalent way to calculate the $Z_{2}$
numbers of a band insulator, which satisfies the following conditions:
first it should use only the periodic bulk system; second, it should
not require any gauge fixing condition - thereby greatly simplifying
the calculation; third, it should be easily applied to general systems
lacking spacial inversion symmetry.

In the present paper we propose a new equivalent expression for the
$Z_{2}$ topological invariant using the U(2N) non-Abelian Berry connection.
Based on this new expression, we further propose a new numerical method
to calculate the $Z_{2}$ topological number for general band insulators,
without choosing a gauge fixing condition. The main idea of the method
is to calculate the evolution of the Wannier function center directly
during a \textquotedbl{}time reversal pumping\textquotedbl{} process,
which is a $Z_{2}$ analog to the charge polarization\cite{king-smith_theory_1993,thouless_quantization_1983}.
We derive that the center of the Wannier function for the effective
1D system can be expressed as the U(1) phase of eigenvalues of a matrix
obtained as the product of the U(2N) Berry connection along the \textquotedbl{}Wilson
loop\textquotedbl{}. The $Z_{2}$ topological numbers can be expressed
as the number of times mod $2$ of the partner switching of these
phases during a complete period of the \textquotedbl{}time reversal
pumping\textquotedbl{} process. Using this new method, we have recalculated
the $Z_{2}$ topological numbers for several TI systems, including
strained HgTe, Bi, Sb and Bi$_{2}$Se$_{3}$, and found the \textquotedbl{}partner
switching\textquotedbl{} patterns, which differentiate between topologically
trivial and nontrivial behavior. The rest of paper will be organized
as follows: in section II we derive the new mathematical form of the
$Z_{2}$ numbers through the \textquotedbl{}Wilson loop\textquotedbl{};
we apply the new method to various TI systems in section III; we prove
the equivalence of the new methods and the $Z_{2}$ number propose
by Fu and Kane\cite{fu_time_2006} in the appendices.

\section{The formalism}

The Bloch wave functions describing the band structure of a translationally
invariant system can be expressed in terms of a complete set of local
basis labeled by the unit cell $i$ and some other quantum number
$\alpha$ as:

\begin{equation}
\left|\psi_{nk}(\mathbf{r})\right\rangle =\frac{1}{\sqrt{N}}\sum_{\alpha i}u_{n\alpha}(\mathbf{k})e^{i\mathbf{k}\cdot\mathbf{R}_{i}}\left|\phi_{\alpha}
(\mathbf{r}-\mathbf{R}_{i})\right\rangle \label{eq:bloch}
\end{equation}

\noindent We will first focus on the 2D system. The topological nature
of a 3D insulator can be determined by looking at 
these effective 2D systems with fixed $k_{i}=0$ and $k_{i}=\pi$
($i=x,y,z$).

Following Fu and Kane, for a 2D insulator we can further reduce the
dimension by fixing $k_{y}$ and study the {}``time reversal'' pumping
process with the adiabatic change of $k_{y}$. The main idea of our
new formalism is to directly look at the evolution of Wannier centers
for these effective 1D systems in the subspace which contains only
occupied states. Fixing $k_{y}$, we can define the position operator
for the effective 1D system as

\begin{equation}
\hat{X}=\sum_{i\alpha}e^{i\delta k_{x}\cdot\mathbf{R}_{i}}|\phi_{\alpha}(\mathbf{r}-\mathbf{R}_{i})
\rangle\langle\phi_{\alpha}(\mathbf{r}-\mathbf{R}_{i})|\label{eq:R_op_1}
\end{equation}

\noindent where $\delta k_{x}\equiv\frac{2\pi}{N_{x}a_{x}}$, $N_{x}$
is the number of real-space unit cells along the $x$ direction, $a_{x}$
is the lattice constant, $\alpha$ is the orbital and spin index and
$\mathbf{R}_{i}$ labels the unit cell. The operator $\hat{X}$ is
a unitary operator with all the eigenvalues being $e^{i\delta k_{x}\cdot\mathbf{R}_{i}}$,
whose phase represents the position. The eigenvalue of the position
operator can be viewed as the center of maximum localized Wannier
function (MLWF) formed by the bands included in the operator $\hat{X}$.
Because the local basis set $\alpha$ is assumed to be complete ,
such MLWFs are always well defined.As pointed out by Fu and Kane,
the $Z_{2}$ topological invariant can be determined by looking at
the evolution of the Wannier function center for the effective 1D
system with fixed $k_{y}$ in the subspace spanned by the occupied
bands only . Therefore we should consider the eigenvalue of the projected
position operator defined as

\begin{align}
\hat{X}_{P} & =\hat{P}\hat{X}\hat{P}\nonumber \\
 & =\sum_{nm\in o}\sum_{k_{x}k_{x}^{\prime},i\alpha}e^{i\delta k_{x}\cdot\mathbf{R}_{i}}|\psi_{nk_{x}k_{y}}\rangle\langle\psi_{nk_{x}k_{y}}|\nonumber \\
 & \times|\phi_{\alpha}(\mathbf{r}-\mathbf{R}_{i})\rangle\langle\phi_{\alpha}
 (\mathbf{r}-\mathbf{R}_{i})|\psi_{mk_{x}^{\prime}k_{y}}
 \rangle\langle\psi_{mk_{x}^{\prime}k_{y}}|\nonumber \\
 & =\sum_{nm\in o}\sum_{k_{x}k_{x}^{\prime},i}e^{i(k_{x}+\delta k_{x}-k_{x}^{\prime})\cdot\mathbf{R}_{i}}|\psi_{nk_{x}k_{y}}
 \rangle\langle\psi_{mk_{x}^{\prime}k_{y}}|\nonumber \\
 & \times\left[\sum_{\alpha}u_{n\alpha}^{*}(k_{x})u_{m\alpha}
 (k_{x}^{\prime})\right]\nonumber \\
 & =\sum_{k_{x}k_{x}^{\prime}}\delta(k_{x}+\delta k_{x}-k_{x}^{\prime})\sum_{nm\in o}|\psi_{nk_{x}k_{y}}\rangle\langle\psi_{mk_{x}^{\prime}k_{y}}|\nonumber \\
 & \times\left[\sum_{\alpha}u_{n\alpha}^{*}(k_{x})u_{m\alpha}
 (k_{x}^{\prime})\right]\label{eq:Xp}
\end{align}

\noindent where $o$ means the occupied bands. The above operator
can be written in a more suggestive matrix form

\begin{equation}
\hat{X}_{P}(k_{y})=\left[\begin{array}{cccccc}
0 & F_{0,1} & 0 & 0 & 0 & 0\\
0 & 0 & F_{1.2} & 0 & 0 & 0\\
0 & 0 & 0 & F_{2,3} & 0 & 0\\
0 & 0 & 0 & 0 & \cdots & 0\\
0 & 0 & 0 & 0 & 0 & F_{N_{x}-2,N_{x}-1}\\
F_{N_{x}-1,0} & 0 & 0 & 0 & 0 & 0\end{array}\right]\label{eq:Xp_mtrx}
\end{equation}

\noindent where $F_{i,i+1}^{nm}(k_{y})=\sum_{\alpha}u_{n\alpha}^{*}(k_{x,i},k_{y})u_{m\alpha}(k_{x,i+1},k_{y})$
are the $2M\times2M$ matrices spanned in $2M$ occupied states and
$k_{x,i}=\frac{2\pi i}{N_{x}a_{x}}$ are the discrete k points taken
along the x-axis.

The eigen problem of $\hat{X}_{P}(k_{y})$ can be solved by the transfer
matrix method. We can define a product of $F_{i,i+1}$ as

\begin{equation}
D(k_{y})=F_{0,1}F_{1,2}F_{2,3}\cdots F_{N_{x}-2,N_{x}-1}F_{N_{x}-1,0}\label{eq:D_ky}
\end{equation}

\noindent $D(k_{y})$ is a $2M\times2M$ matrix, which has $2M$ eigenvalues:

\[
\lambda_{m}^{D}=|\lambda_{m}^{D}|e^{i\theta_{m}^{D}}\begin{array}{cccc}
\, & \, & \, & \,\end{array}m=1,2,\cdots,2M\]

\noindent where $\theta_{m}^{D}$ is the phase of the eigenvalues:

\begin{equation}
\theta_{m}^{D}=Im(log\lambda_{m}^{D})\label{eq:argument}
\end{equation}

We can easily prove that the eigenvalue of projected position operator
$X_{P}$ can be simply related to the eigenvalue of the above D-matrix
by

\begin{equation}
\lambda_{m,n}^{P}=\sqrt[N_{x}]{\lambda_{m}^{D}}=\sqrt[N_{x}]
{|\lambda_{m}^{D}|}e^{i(\theta_{m}^{D}+2\pi n)/N_{x}}\label{eq:lambda}
\end{equation}

\noindent where $n=1,2,\cdots,N_{x}$. We can further prove that the
D-matrix is unitary and all the $|\lambda_{m}^{D}|$ equals one.

\begin{figure}[h]

\begin{centering}
\includegraphics[scale=0.25,angle=-90]{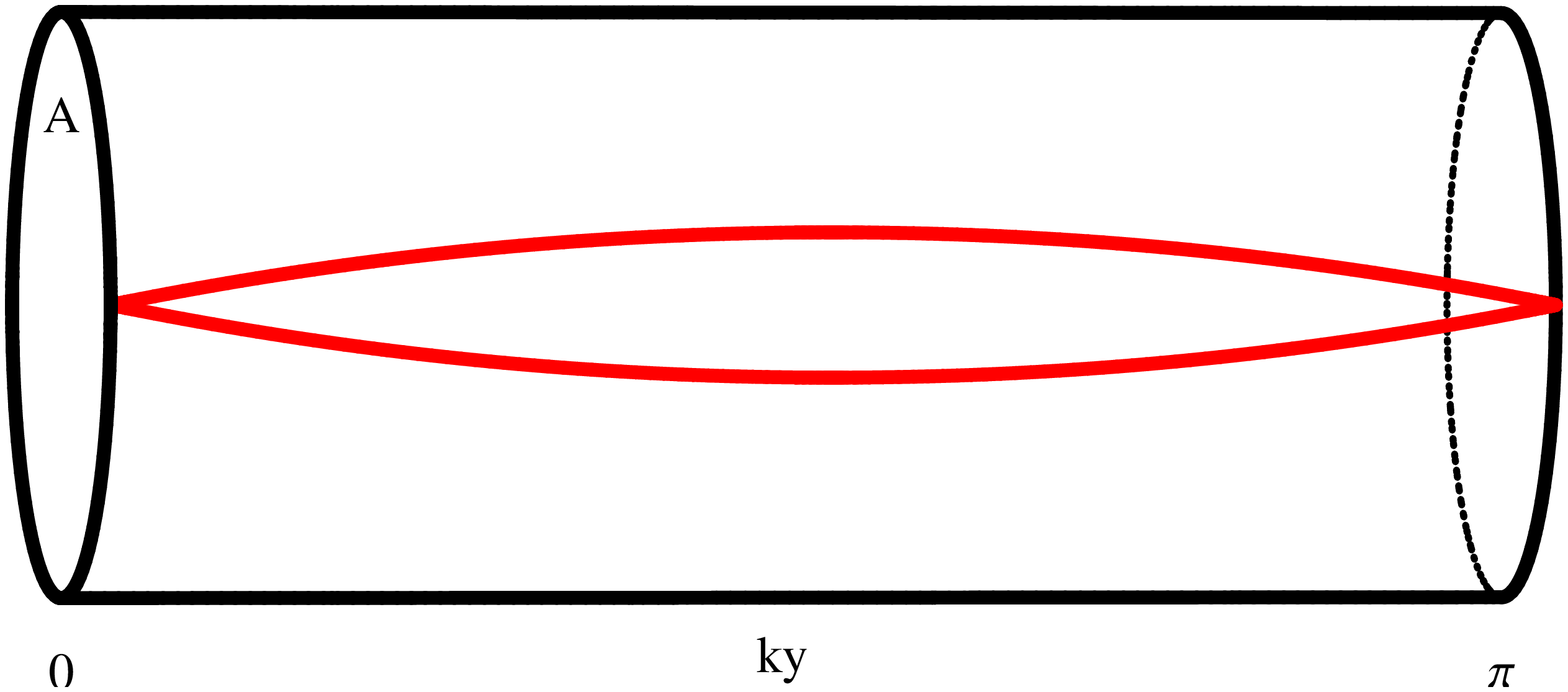}
\par\end{centering}

\begin{centering}
\includegraphics[scale=0.25,angle=-90]{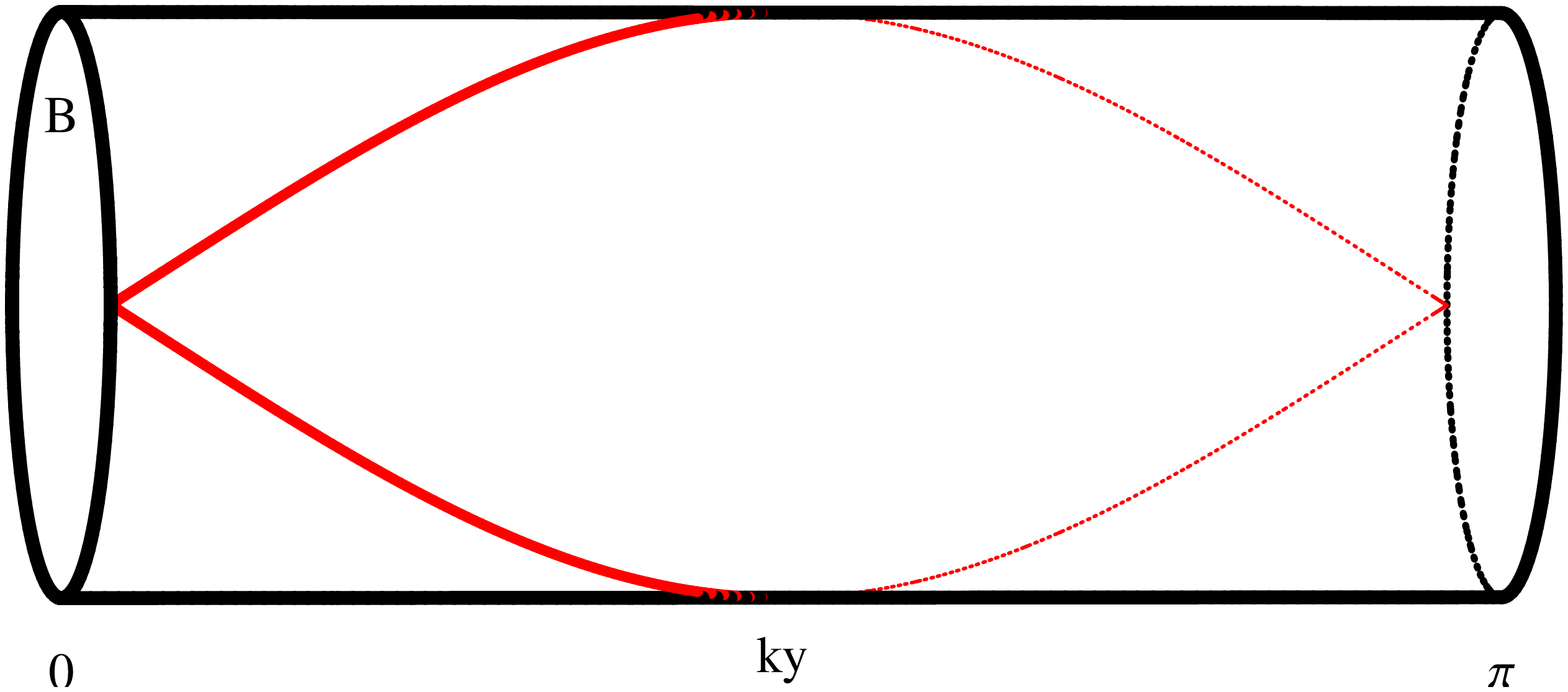}
\par\end{centering}

\begin{centering}
\includegraphics[scale=0.25,angle=-90]{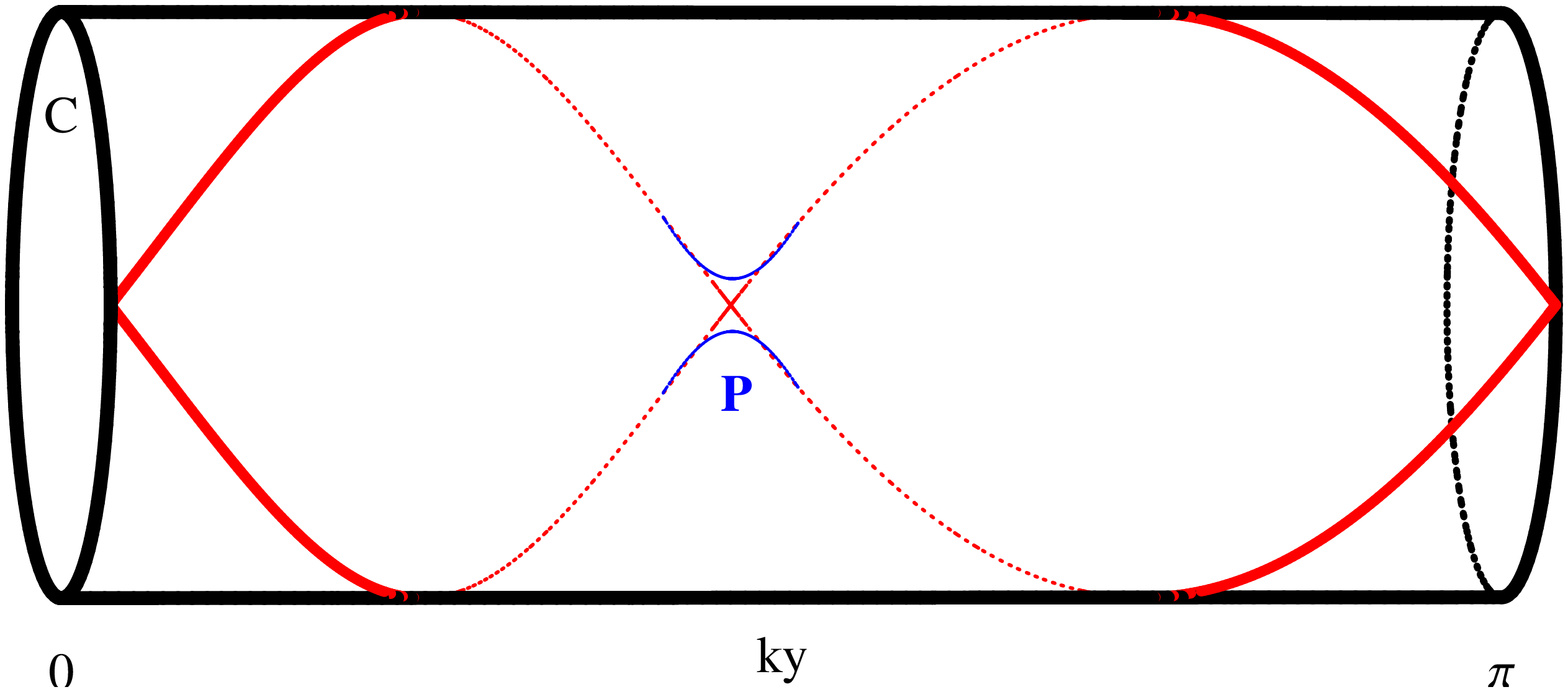}
\par\end{centering}

\begin{centering}
\includegraphics[scale=0.25,angle=-90]{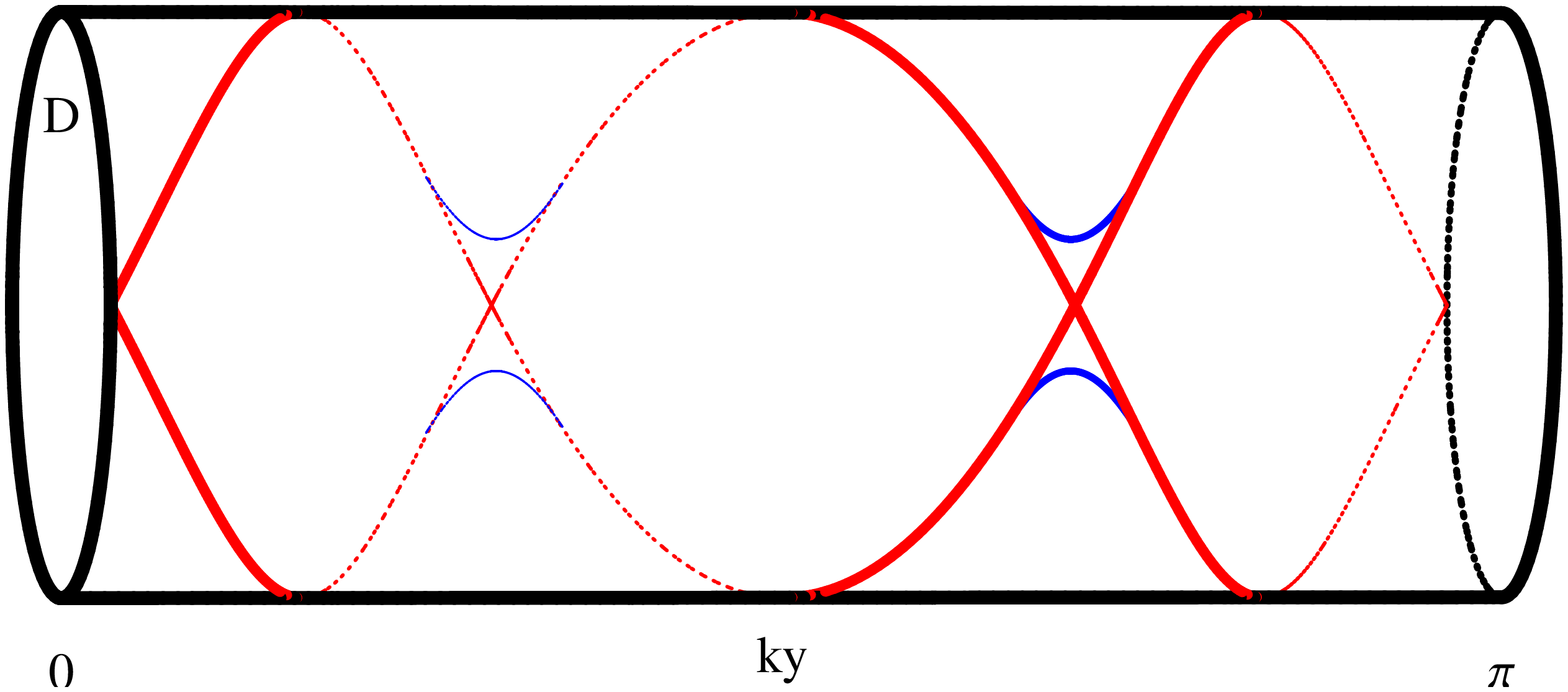}
\par\end{centering}

\caption{Schematic plots of the Wannier center curves: (A) for the trivial
case, the Wannier center winding the cylinder zero times; (B) the
Wannier center winding the cylinder one times; (C) the Wannier center
winding the cylinder twice and the cross point labeled by $P$ is
not protected by time reversal symmetry and it is usually eliminated
by some perturbation terms. (D)the Wannier center winding the cylinder
3 times, which is topologically equal to the case in B. 
\label{fig:schematic}}

\end{figure}

The evolution of the Wannier function center for the effective 1D
system with $k_{y}$ can be easily obtained by looking at the phase
factor $\theta_{m}^{D}$. At $k_{y}=0$, the eigenvalues of the D-matrix
appear in degenerate pairs due to time reversal symmetry, which results
in pairs of Wannier centers sitting at $k_{y}=0$. When $k_{y}$ moves
away from the origin, the Wannier center pairs split and recombine
at $k_{y}=\pi$, as shown in Fig.\ref{fig:schematic}. Because $\theta_{m}^{D}$
is a phase factor, when two $\theta_{m}^{D}$s meet together, they
may differ by integer times of $2\pi$. Therefore the evolution of
each Wannier center pair will enclose the whole cylinder an integer
times, which can be viewed as the winding number of the Wannier center
pair. The Z$_{2}$ topological number is related to the summation
of the winding numbers for all the pairs. If it is odd, then the Z$_{2}$
topological number is odd. It seems that the total winding number
of the Wannier center pairs should generate an integer class Z instead
of Z$_{2}$. To clarify this point, let's look at the evolution of
a Wannier center pair with winding number $4\pi$. In that particular
case, as shown in Fig.\ref{fig:schematic}(C), the pair of Wannier
centers must have an extra \textquotedbl{}accidental\textquotedbl{}
degeneracy between $k_{y}=0$ and $\pi$, which is not protected by
any symmetry and can be removed by \textquotedbl{}deforming\textquotedbl{}
the Hamiltonian slightly to make the crossing of the levels become
an anti-crossing as shown in Fig.\ref{fig:schematic}(C). The \textquotedbl{}deformation\textquotedbl{}
process will thus change the total winding number by $4\pi$ and make
it 0. Therefore only the total winding number mod $2$ is a topological
invariant.

The eq.(\ref{eq:D_ky}) can be viewed as the discrete expression of
the Wilson loop for the U(2M) non-Abelian Berry's connection. It is
obviously invariant under the U(2M) gauge transformation and thus
can be calculated directly from the wave functions obtained by first
principle method without choosing any gauge fixing condition, which
is the biggest advantage of the present form of the $Z_{2}$ invariance.
The equivalence to the $Z_{2}$ number proposed by Kane and Mele will
be proved rigorously in the appendix.

\section{Numerical results}

In the present section, we will implement the method and explicitly
compute the $Z_{2}$ invariant for a series of systems. For each particular
system, we calculate the evolution of the $\theta^{D}$ defined in
equation(\ref{eq:argument}) as the function of $k_{y}$ from zero
to $\pi$. The winding number of the Wannier center pairs defined
in the above section can be checked in an equivalent way which is
much simple in practice. We first draw an arbitrary reference line
parallel to the $k_{y}$ axis, then compute the $Z_{2}$ number by
counting how many times the evolution lines of the Wannier centers
crosses the reference line.

\subsection{BHZ model }

Bernevig Hughes and Zhang (BHZ) showed that for an appropriate range
of well thickness, the HgTe/CdTe quantum well exhibits an inverted
sub-band structure. In this inverted regime, the system exhibits a
2D quantum spin Hall effect\cite{bernevig_quantum_2006-1}. BHZ introduce
a simple four band tight binding model to describe this effect:

\begin{equation}
H_{eff}(k_{x},k_{y})=\left[\begin{array}{cc}
H(\mathbf{k}) & 0\\
0 & H^{*}(-\mathbf{k})\end{array}\right]\label{eq:H_HgTe}\end{equation}

\noindent where $H(\mathbf{k})=\varepsilon(\mathbf{k})+d_{i}(\mathbf{k})\sigma_{i}$,
$d_{1}+id_{2}=A[\sin k_{x}+i\sin k_{y}]$,$d_{3}=-2B[2-\frac{M}{2B}-\cos k_{x}-\cos k_{y}]$,$\varepsilon(\mathbf{k})=C-2D[2-\cos k_{x}-\cos k_{y}]$.
Real HgTe does not have inversion symmetry but the BHZ toy model does.
To describe the inversion symmetry breaking effect we add a new term
$H^{\prime}$: 
\begin{equation}
H^{\prime}=\left[\begin{array}{cccc}
0 & 0 & 0 & \Delta\\
0 & 0 & -\Delta & 0\\
0 & -\Delta & 0 & 0\\
\Delta & 0 & 0 & 0\end{array}\right]\label{eq:H_HgTe_imp}
\end{equation}

\begin{figure}[h]

\begin{centering}
\includegraphics[scale=0.65,angle=-90]{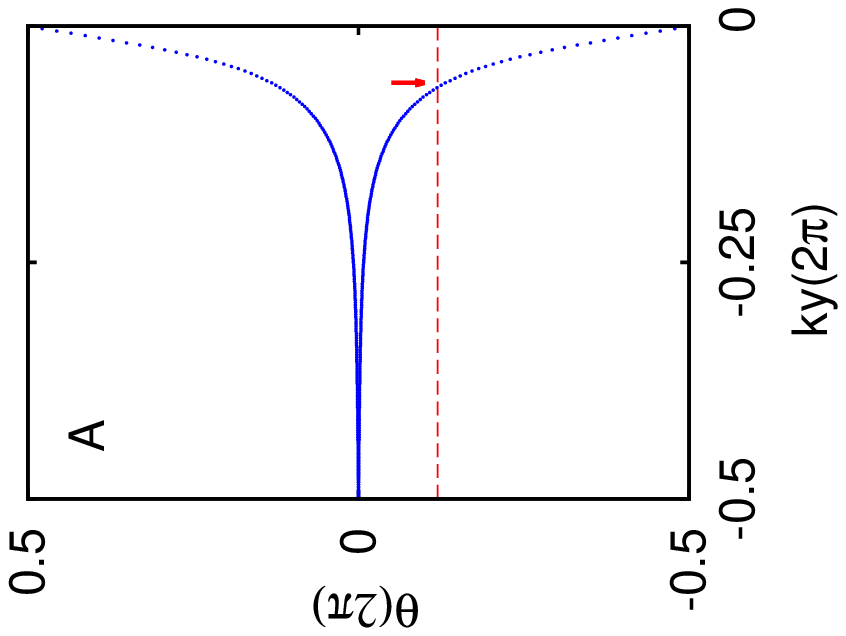}
\includegraphics[scale=0.65,angle=-90]{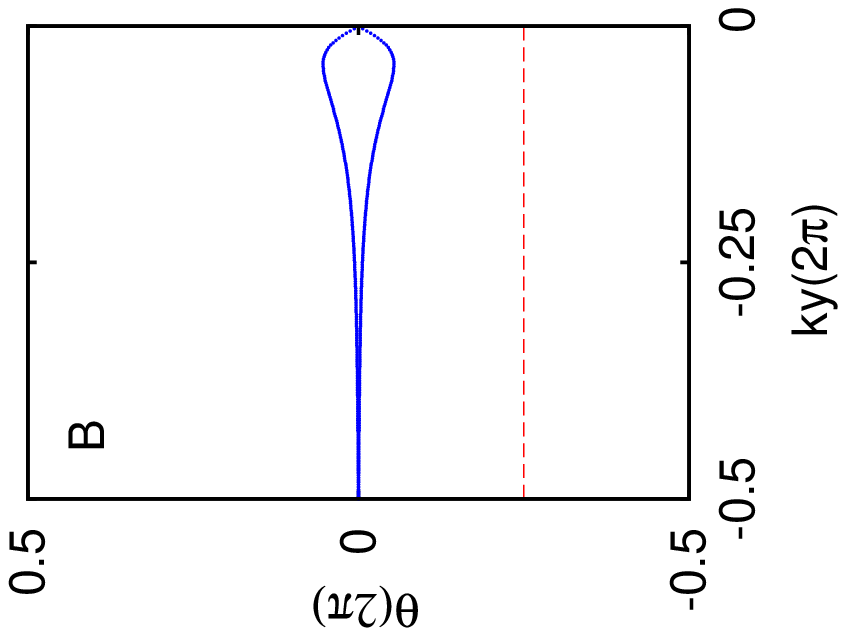}
\par\end{centering}

\caption{Wannier centers for the BHZ model. (A) For the QSH phase (A=-13.68,
B=-16.9, C=-0.0263, D=-0.514, M=-2.058, $\Delta$=1.20) , The Wannier
center cross the reference line (red dashed line) once (odd times);
(B) For the Normal insulating phase (A=-14.48, B=-18.0, C=-0.018,
D=-0.594, M=2.766, $\Delta$=1.20), The Wannier center cross the reference
line (red dashed line) zero (even) times. \label{fig:BHZ}}

\end{figure}

\noindent We apply the new method to calculate the shift of the Wannier
function center based on the above model Hamiltonian and show the
results in Fig.\ref{fig:BHZ}.

To show these results more clearly, we glue $\theta=-\pi$ line and
$\theta=\pi$ line together. Then the Wannier centers live on a cylinder
surface. The corresponding results for the QSH phase are showed in
Fig.\ref{fig:BHZ}(A). When moving from $k_{y}=0$ to $\pi$ we see
that the two evolution lines of Wannier centers enclose the cylinder
once and equivalently these evolution lines cross the reference line
(the red dashed line) only once (odd times). By contrast, for the
normal insulator phase, as shown in Fig.\ref{fig:BHZ}(B), the two
evolution lines of Wannier centers never cross the reference line.
Therefore in the BHZ toy model for TI, the $Z_{2}$ number calculated
by our new method is consistent with the previous conclusion. Next
we will apply the method to more realistic models of insulating materials.

\subsection{\textcolor{black}{CdTe and HgTe} }

\begin{figure}[h]
\begin{centering}
\includegraphics[scale=0.65,angle=270]{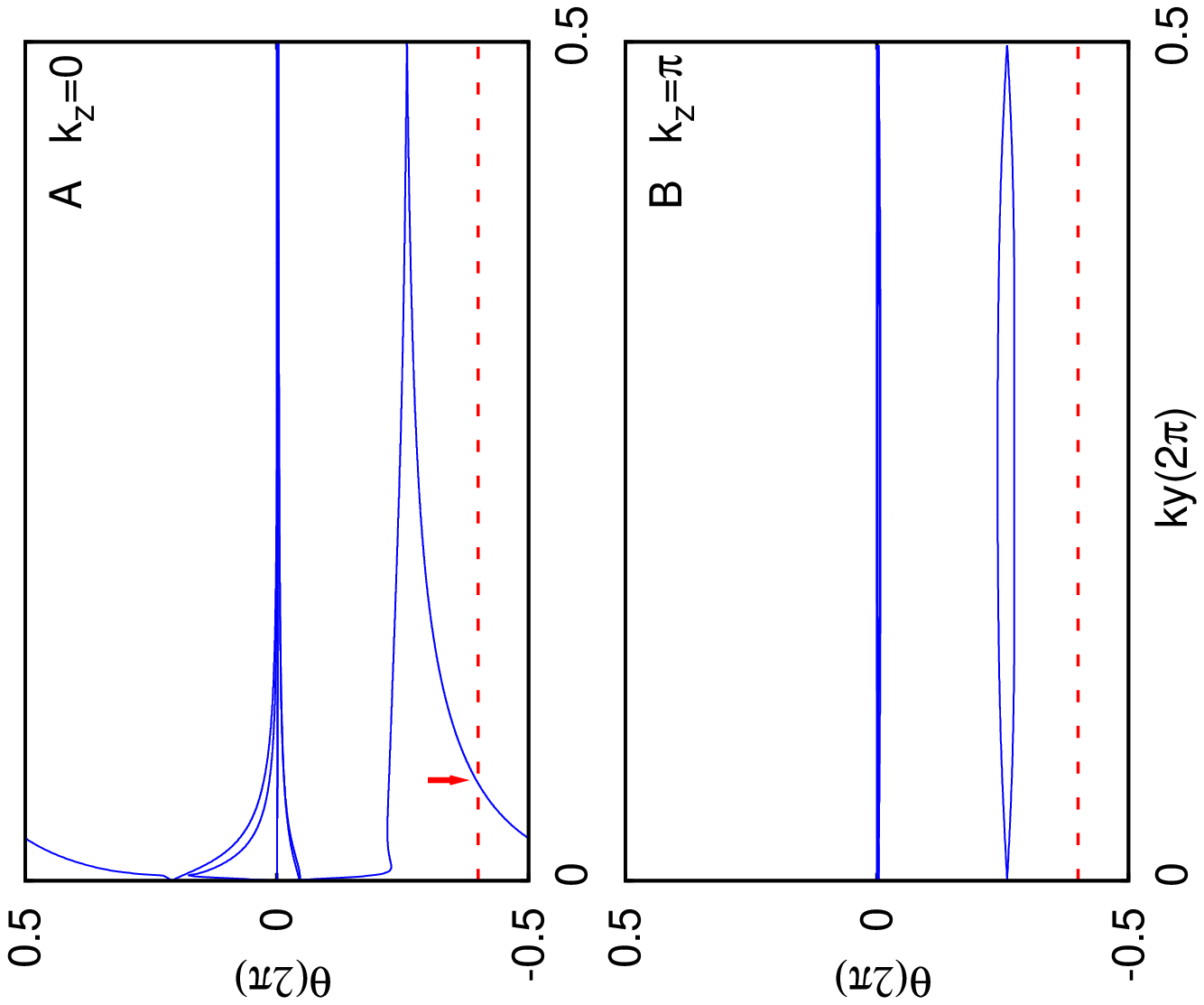}
\par\end{centering}

\begin{centering}
\includegraphics[scale=0.65,angle=270]{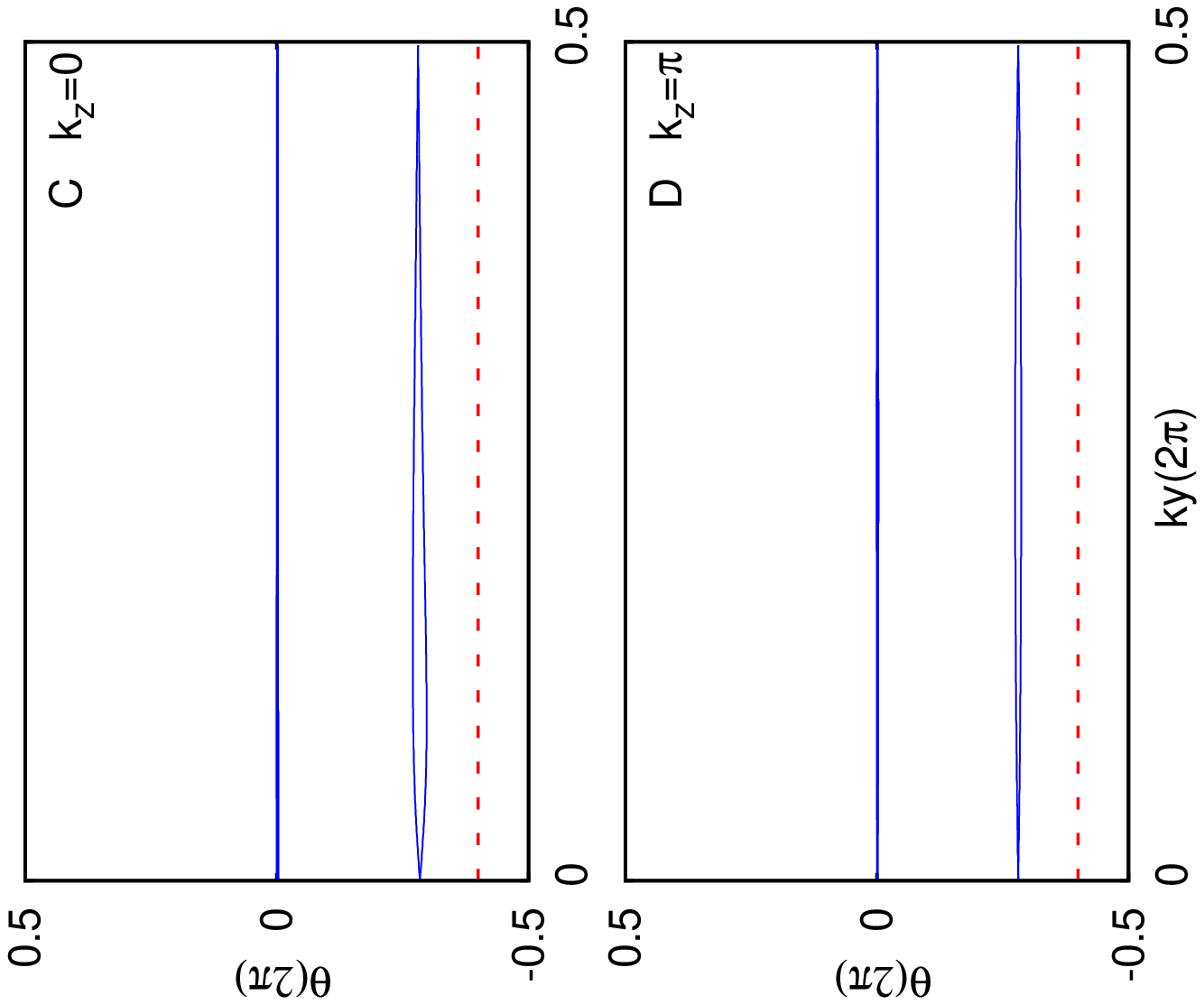}
\par\end{centering}

\caption{\label{fig:HgTe_CdTe} The evolution lines of Wannier centers for
HgTe (A, B) and CdTe (C, D). For HgTe system, The Wannier center cross
the reference line odd times in the $k_{z}=0$ plane and even times
in the $k_{z}=\pi$ plane, indicating HgTe is a strong topological
insulator. For CdTe, the Wannier centers cross the reference line
zero times for both $k_{z}=0$ and $\pi$ planes, indicating CdTe
is a normal insulator.}

\end{figure}

The CdTe and HgTe materials have a similar zinc-blende structure without
bulk inversion symmetry. CdTe has an normal electronic structure,
where the conduction bands ($\Gamma_{6}$) have the s-like character
and the valance bands have the p-like character($\Gamma_{8}$) through
out the whole Brillouin Zone. In HgTe, the band structure is inverted
in a small area near the $\Gamma$ point, where the s-like $\Gamma_{6}$
band sinks below the p-like $\Gamma_{8}$ band. The band inversion
at the $\Gamma$ point changes the topological nature of the band
structure and makes the HgTe to be the topological insulator if a
true energy gap is opened by the lattice distortion\cite{daixi_helical_2008}(As
pointed out in ref.\onlinecite{daixihelical2008} the uniaxial strain
is applied along the {[}001{]} direction for HgTe by choosing the
$c/a$ ratio to be 0.98 and the energy gap is about 0.05eV at the
$\Gamma$ point). We then apply a tight-binding model\cite{kobayashi_chemical_1982}
to calculate the pattern of the Wannier center evolution $\theta$
defined in Eq.\ref{eq:argument} and show the results in Fig.\ref{fig:HgTe_CdTe}.
It is very clear that in the HgTe system, for $k_{z}=0$ the evolution
line crosses the reference line (red dashed line) once (as shown in
Fig.\ref{fig:HgTe_CdTe}(A)), while for $k_{z}=\pi$ it never crosses
(as shown in Fig.\ref{fig:HgTe_CdTe}(B)). The above results indicate
that in the case of HgTe the effective $2D$ systems for fixed $k_{z}=0$
and $\pi$ are effectively 2D topological insulator and normal insulator
respectively, which determines HgTe to be a strong 3D topological
insulator\cite{fu_topological_2007}. A similar analysis can be also
applied to CdTe and the results are shown in Fig.\ref{fig:HgTe_CdTe}(C)
and (D). They clearly indicate that CdTe is a normal insulator.

\subsection{Bi$_{2}$Se$_{3}$ system}

Recently, the tetradymite semiconductors Bi$_{2}$Te$_{3}$, Bi$_{2}$Se$_{3}$,
and Sb$_{2}$Te$_{3}$ have been theoretically predicted and experimentally
observed to be topological insulators (TI) with a bulk band gap as
large as 0.3eV in Bi$_{2}$Se$_{3}$\cite{zhang_topological_2009,xia_observation_2009,
chen_experimental_2009,hsieh_observation_2009-1,hsieh_tunable_2009,park_quasiparticle_2010}.
The Bi$_{2}$Se$_{3}$ surface state has been found by both ARPES\cite{xia_observation_2009,chen_experimental_2009}
and STM \cite{zhang_experimental_2009}, consistent with the theoretical
results\cite{zhang_topological_2009}.

\begin{figure}[H]

\begin{centering}
\includegraphics[scale=0.65,angle=-90]{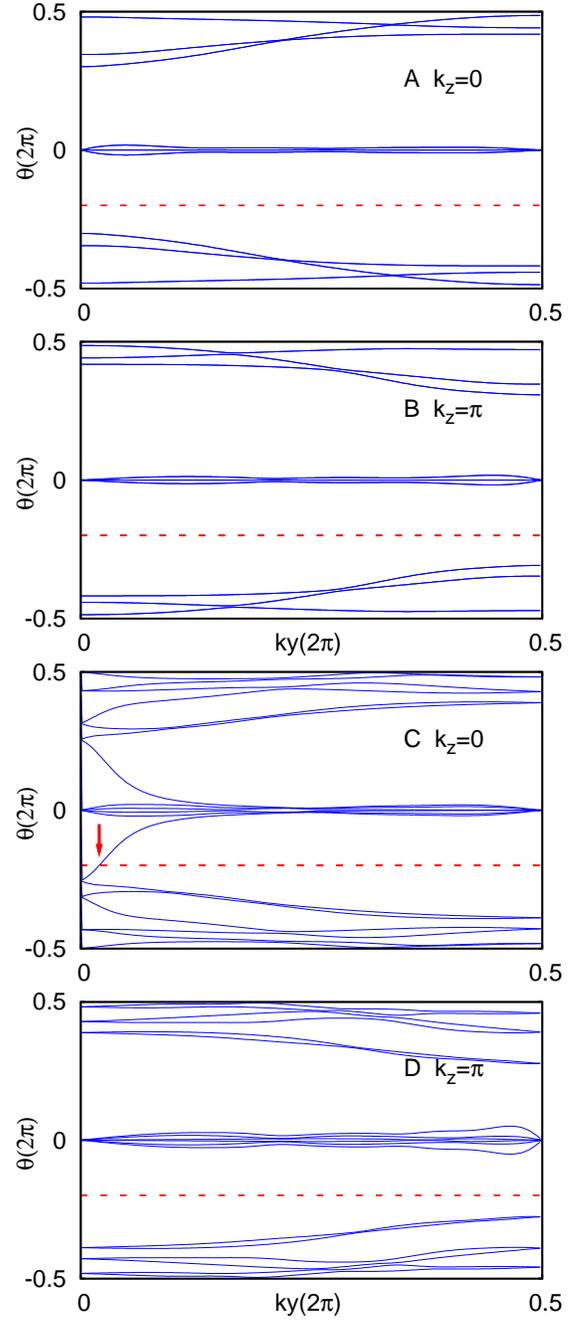}
\par\end{centering}

\caption{\label{fig:BiSe} The evolution lines of Wannier centers for Bi$_{2}$Se$_{3}$
system without(A, B) and with SOC (C, D). If we turn off SOC the system
is a normal insulator and the Wannier centers never cross the reference
line for both $k_{z}=0$ and $\pi$ planes as shown in (A) and (B),
indicating it is a normal insulator. When the SOC is turned on the
system is in a strong topological phase, the Wannier centers cross
the reference line odd times in $k_{z}=0$ and even times in $k_{z}=\pi$
plane.}

as shown in (C) and (D), indicating the system is topologically non-trivial.
\end{figure}

Since the Bi$_{2}$Se$_{3}$ family has inversion symmetry, the $Z_{2}$
topological number can be easily calculated by the product of half
the parities at each high symmetry points in the Brillouin Zone\cite{fu_topological_2007-1}.
Below we apply our new method to calculate the topological property
of this system, using the tight binding model based on the Wannier
functions obtained in reference\cite{zhang_topological_2009}. We
first perform the calculation for the Bi$_{2}$Se$_{3}$ without spin-orbit
coupling: the results are shown in Fig.\ref{fig:BiSe}(A,B). It is
clear that the evolution lines never cross the reference line for
both $k_{z}=0$ and $\pi$, indicating that the system is topologically
trivial without spin-orbital coupling. When the realistic spin orbital
coupling is turned on, as shown in Fig.\ref{fig:BiSe}(C,D), the evolution
lines cross the reference line once only in the case of $k_{z}=0$
but not for $k_{z}=\pi$ indicating the Bi$_{2}$Se$_{3}$ bulk material
is a 3D strong topological insulator.

\subsection{Bi$_{2}$Te$_{3}$ slab system}

\begin{figure}[H]

\begin{centering}
\includegraphics[scale=0.65,angle=-90]{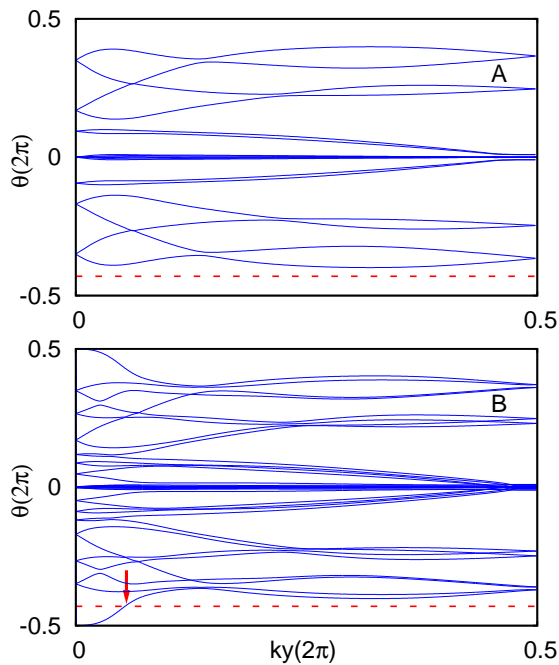}
\par\end{centering}

\caption{\label{fig:bite_slab} The evolution lines of Wannier centers for
1QL and 2QL Bi$_{2}$Te$_{3}$ slab. (A) The 1QL Bi$_{2}$Te$_{3}$
slab is in normal insulator phase. (B) The 2QL Bi$_{2}$Te$_{3}$
slab is in topological insulator phase.}

\end{figure}

As calculated by Liu et al\cite{liu_oscillatory_2010}, upon reducing
the thickness of Bi$_{2}$Te$_{3}$ and Bi$_{2}$Se$_{3}$ films,
the topological nature of the system alternates between topologically
trivial and non-trivial behavior as a function of the layer thickness.
Liu et al. point out that the 1QL Bi$_{2}$Te$_{3}$ slab is a trivial
insulator and 2QL Bi$_{2}$Te$_{3}$ slab is a 2D topological insulator\cite{liu_oscillatory_2010}.
We apply our method to these systems. The evolution patterns for the
1QL and 2QL Bi$_{2}$Te$_{3}$ slabs are obtained using the tight
binding Hamiltonian developed in references\cite{liu_oscillatory_2010,zhangwei_first-principles_2010}
and the results are summarized in Fig.\ref{fig:bite_slab}. In the
1QL slab system (Fig.\ref{fig:bite_slab}(A)), the evolution pattern
appears in a trivial manner while that of the 2QL slab system is non-trivial(Fig.\ref{fig:bite_slab}(B)).
This is consistent with the conclusion based on the parity counting\cite{liu_oscillatory_2010}.

\subsection{Bi and Sb system }

\begin{figure}[H]

\begin{centering}
\includegraphics[scale=0.65,angle=270]{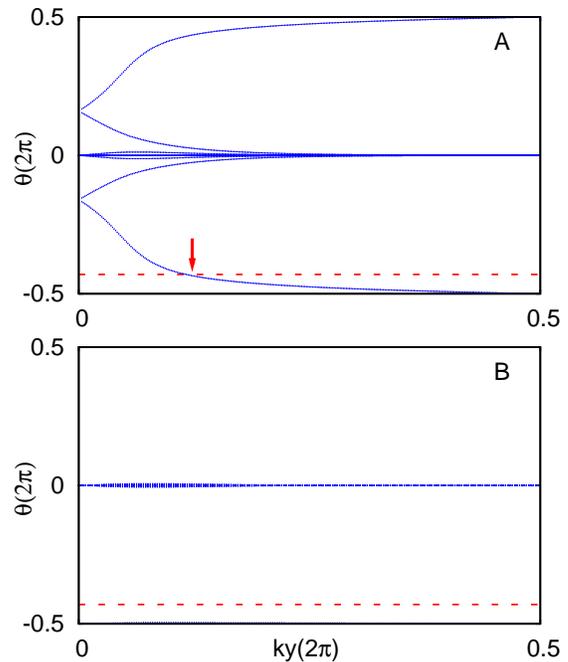}
\par\end{centering}

\caption{\label{fig:Bi_and_Sb} The evolution lines of Wannier centers for
2D single-bilayer Bi (A) and Sb (B) system, indicating the single-bilayer
Bi is topologically nontrivial but Sb is topologically trivial.}

\end{figure}

Murakami pointed out that the $Z_{2}$ topological number is odd in
the 2D bilayer bismuth system \cite{murakami_quantum_2006}. We apply
our method to this system using the tight-binding model developed
in reference\cite{liu_electronic_1995}, which faithfully reproduces
the bulk bismuth band structure. As shown in Fig.\ref{fig:Bi_and_Sb}A,
the band structure of bilayer bismuth is topologically nontrivial,
which is quite consistent with the previous conclusion \cite{murakami_quantum_2006}.
After that we apply the same method to calculate single-bilayer Sb,
which has the similar lattice structure as bismuth, but with relatively
weak SOC. As plotted in Fig.\ref{fig:Bi_and_Sb}(B), the evolution
pattern of bilayer Sb shows clearly that it is in the normal insulator
phase, which is also consistent with the parity counting.

\subsection{Graphene system }

\begin{figure}[H]

\begin{centering}
\includegraphics[scale=0.65,angle=270]{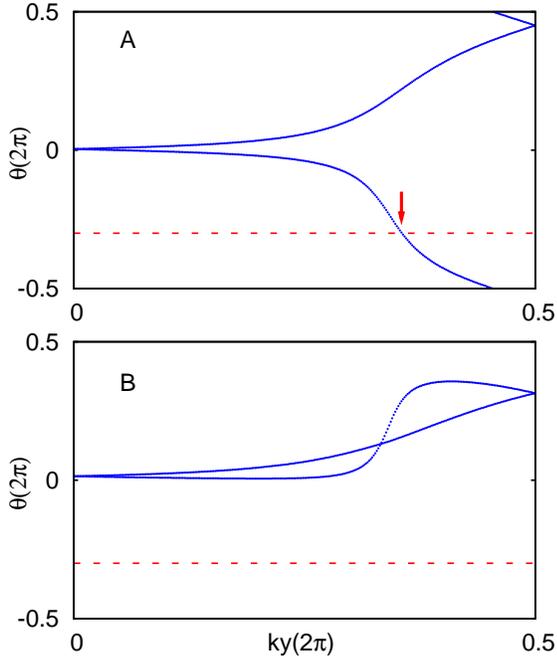}
\par\end{centering}

\caption{\label{fig:graphene} The evolution lines of Wannier center for graphene
in the (A) QSH phase $\lambda_{\upsilon}=0.1t$ and (B) the normal
insulating phase $\lambda_{\upsilon}=0.4t$. In both cases $\lambda_{SO}=0.06t$
and $\lambda_{R}=0.05t$.}
\end{figure}

In 2005 Kane and Mele pointed out \cite{kane_z_2_2005} that there
are two different phases in graphene, depending on the spin-orbital
coupling $\lambda_{R}$ and the staggered sublattice potential parameter
$\lambda_{v}$. The system is in a quantum spin Hall (QSH) phase when
$\lambda_{v}=0.1t$ and normal insulator phase when $\lambda_{v}=0.4t$,
where $t$ is the hopping parameter. In the present study, we use
the same parameters as Kane and Mele to calculate the corresponding
Wannier center evolution pattern. The results are shown in Fig.\ref{fig:graphene}.
It can be easily found that there is one partner switching in Fig.\ref{fig:graphene}(A)
but not in Fig.\ref{fig:graphene}(B) indicating the former is topologically
non trivial and the latter is trivial.

In conclusion, we have proposed a new equivalent expression for the
$Z_{2}$ topological invariance using the $U(2N)$ non-Abelian Berry
connection. Based on this new expression we calculated the evolution
of the Wannier function center for several topological and normal
insulating systems with or without inversion symmetry. We showed that
for the nontrivial topological insulators, the Wannier function center
have partner switching patterns, topologically different from the
normal (trivial) insulating systems. Additionally, we gave a proof
that the new method is equivalent to the $Z_{2}$ number proposed
by Fu and Kane.

\emph{Note} During the preparation of this manuscript, we noticed
the paper by Soluyanov and Vanderbilt\cite{soluyanov_wannier_2010},
where the construction of Wannier functions for $Z_{2}$ topological
insulators are discussed from a different view of point. We addressed
in this paper that the real construction of Wannier functions is not
necessary, while only the \textquotedbl{}Wannier representation\textquotedbl{}
and corresponding Berry connection evaluated along the \textquotedbl{}Wilson
loop\textquotedbl{} are essential keys in order to identify the topological
nature.

\section*{Acknowledgments}

BAB was supported by Princeton Startup Funds, Alfred P. Sloan Foundation,
NSF DMR- 095242, and NSF China 11050110420, and MRSEC grant at Princeton
University, NSF DMR-0819860. XLQ is partly supported by Alfred P.
Sloan Foundation. BAB and XLQ thank the Institute of Physics in Beijing,
China for generous hosting. XD and ZF acknowledge the supports from
NSF of China and that from the 973 program of China (No.2007CB925000)

\appendix

\section{Proof of gauge invariance of the $D(k_{y})$ operator and its equivalence
to the $U(2)$ wilson loop}

We first give the proof of gauge invariance of the $D(k_{y})$ operator
and its equivalence to the $U(2)$ Wilson loop. The $D(k_{y})$ operator
is, at each $k_{y}$, the product:

\begin{equation}
D(k_{y})=F_{0,1}F_{1,2}F_{2,3}F_{N_{x}-2,N_{x}-1}F_{N_{x}-1,0}\label{eq:D}
\end{equation}

\noindent where

\begin{equation}
F_{i,i+1}^{mn}=\langle m,k_{x,i},k_{y}|n,k_{x,i+1},k_{y}\rangle
\end{equation}

\noindent $k_{x,i}=i\Delta k$, $\Delta k=\frac{2\pi}{N_{x}}$, and
let $k_{x,N_{x}}=0$ to make $k_{x}$ go back to the initial point.
$m,n$ are running through from 1 to the number of occupied bands,
$|n,\vec{k}\rangle$ is the n'th wavevector of energy $E_{n}$ of
the insulating Hamiltonian:

\begin{equation}
H(\vec{k})|n,\vec{k}\rangle=E_{n}|n,\vec{k}\rangle\label{eq:H}
\end{equation}

\noindent For a fixed $k_{y}$, the $k_{x}=[-\pi,\pi]$ interval is
broken up in $N_{x}$ parts. This corresponds to periodic boundary
conditions in the $x$ direction and $N_{x}$ sites. $D(k_{y})$ operator
is then:

\begin{align}
D_{mn}(k_{y}) & =[F_{0,1}F_{1,2}F_{2,3}F_{N_{x}-2,N_{x}-1}F_{N_{x}-1,0}]_{mn}\nonumber \\
 & =\langle m,k_{0}|n_{1},k_{1}\rangle\langle n_{1},k_{1}|n_{2},k_{2}\rangle\cdots\nonumber \\
 & \times\langle n_{N_{x}-2},k_{N_{x}-2}|n_{N_{x}-1},k_{N_{x}-1}\rangle\langle n_{N_{x}-1},k_{N_{x}-1}|n,k_{0}\rangle
 \end{align}

\noindent where double index of $n_{i}$ implies summation. As the
$k_{y}$ index is self-understood, we will suppress it at times and
re-introduce it when necessary.

Under a unitary transformation $|n,\vec{k}\rangle$ become to :

\[
|n,\vec{k}\rangle\rightarrow\sum_{m}M_{mn}(\vec{k})|m,\vec{k}\rangle\]

\noindent and hence $D(k_{y})$ matrix goes into\begin{align}
D_{mn}(k_{y}) & \rightarrow M_{m^{\prime}m}^{*}(k_{0})\langle m^{\prime},k_{0}|n_{1}^{\prime},k_{1}\rangle\nonumber \\
 & \times M_{n_{1}^{\prime}n_{1}}(k_{1})\cdots M_{n_{N_{x}-2}^{\prime\prime}n_{N_{x}-2}}^{*}(k_{N_{x}-2})\nonumber \\
 & \times\langle n_{N_{x}-2}^{\prime\prime},k_{N_{x}-2}|n_{N_{x}-1}^{\prime},k_{N_{x}-1}\rangle M_{n_{N_{x}-1}^{\prime}n_{N_{x}-1}}(k_{N_{x}-1})\nonumber \\
 & \times M_{n_{N_{x}-1}^{\prime\prime}n_{N_{x}-1}}^{*}(k_{N_{x}-1})\langle n_{N_{x}-1}^{\prime\prime},k_{N_{x}-1}|n^{\prime},k_{0}\rangle M_{n^{\prime}n}(k_{0})\nonumber \\
= & M_{m^{\prime}m}^{*}(k_{0})\langle m^{\prime},k_{0}|n_{1},k_{1}\rangle\cdots\nonumber \\
 & \times\langle n_{N_{x}-1},k_{N_{x}-1}|n^{\prime},k_{0}\rangle M_{n^{\prime}n}(k_{0})
\end{align}

\noindent or

\[
D(k_{y})\rightarrow M^{\dagger}(k_{0})D(k_{y})M(k_{0})\]

\noindent which means that the eigenvalues of $D(k_{y})$ (or its
trace and determinant) remain unchanged under a $U(occupied)$ gauge
transformation.


For infinitesimal $\Delta k\ll2\pi$, we have:

\begin{align}
F_{i,i+1}^{mn} & =\langle m,k_{x,i},k_{y}|n,k_{x,i+1},k_{y}\rangle\nonumber \\
 & =\delta_{mn}+\langle m,k_{x,i},k_{y}|(|n,k_{x,i+1},k_{y}\rangle-|n,k_{x,i},k_{y}\rangle)\nonumber \\
 & =\delta_{mn}+A_{i,i+1}^{mn}\Delta k\nonumber \\
 & \approx e^{A_{i,i+1}^{mn}\Delta k}\label{eq:F(i,i+1)}
\end{align}

\noindent where $A_{i,i+1}^{mn}=\frac{\langle m,k_{x,i},k_{y}|(|n,k_{x,i},k_{y}\rangle-|n,k_{x,i},k_{y}\rangle)}{\Delta k}$
is the non-abelian $U(occupied)$ gauge field. Hence we have that
$D(k_{y})$ is:

\begin{align}
D(k_{y}) & =\bigg[\prod_{i=0}^{N_{x}-1}F_{i,i+1}\bigg]=\bigg[\prod_{i=0}^{N_{x}-1}e^{A_{i,i+1}\Delta k}\bigg]\nonumber \\
 & =\bigg[Pe^{\int_{C_{ky}}A(k)dk}\bigg]
\end{align}

\noindent This is just the $U(\text{occupied})$ (\emph{not} $SU(\text{occupied})$)
Wilson loop,
where the contour $C_{k_{y}}$is a contour at fixed $k_{y}$ which
goes across the BZ in $k_{x}$, i.e. goes from $k_{x}=-\pi$ to $k_{x}=\pi,$
through $k_{x}=0$.

\section{Wilson loop and pfaffian topological invariant}

In this section, we will prove that the $U(2)$ Wilson loop is related
to the topological invariant of time-reversal topological insulators.
In a system with time-reversal invariant we can relate the bands at
$k$ and $-k$ through a unitary matrix $B$,

\begin{equation}
|n,-k\rangle=B_{nm}^{*}(k)\hat{T}|m,k\rangle\label{eqn:Ttrans}
\end{equation}

\noindent with $B(k)$ unitary and has the property:

\begin{equation}
B(-k)=-B^{T}(k)\label{eq:bk-bmk}
\end{equation}

\noindent We have the following relation between $F_{k_{1},k_{2}}$
matrices:

\begin{align}
F_{-k_{2},-k_{1}}^{mn} & =\langle m,-k_{2}|n,-k_{1}\rangle\nonumber \\
 & =\langle m^{\prime},k_{2}|\hat{T}B_{mm^{\prime}}B_{nn^{\prime}}^{*}\hat{T}|n^{\prime},
 k_{1}\rangle\nonumber \\
 & =B_{mm^{\prime}}B_{nn^{\prime}}^{*}F_{k_{1},k_{2}}^{n^{\prime}m^{\prime}}\nonumber \\
 & =B_{mm^{\prime}}(F_{k_{1},k_{2}}^{T})^{m^{\prime}n^{\prime}}B_{n^{\prime}n}^{\dagger}
\end{align}

\noindent or equivalently:

\begin{equation}
F_{-k_{2},-k_{1}}=B(k_{2})F_{k_{1},k_{2}}^{T}B^{\dagger}(k_{1})\label{eq:F(-k)}
\end{equation}

We now focus on the $k_{y}=0$ or $k_{y}=\pi$ paths (say $k_{y}=0$),
each of which has $k_{x}$ going form $-\pi$ to $\pi$, so that $k_{y}=-k_{y}$,
and compute the finite difference:

\begin{align}
D(k_{y}=0) & =\prod_{i=0}^{N_{x}-1}F_{i,i+1}\nonumber \\
 & =F_{-\frac{N_{x}}{2}\Delta k,-(\frac{N_{x}}{2}-1)\Delta k}\cdots F_{-3\Delta k,-2\Delta k}F_{-2\Delta k,-\Delta k}\nonumber \\
 & \times F_{-\Delta k,0}F_{0,\Delta k}F_{\Delta k,2\Delta k}\cdots F_{(\frac{N_{x}}{2}-1)\Delta k,\frac{N_{x}}{2}\Delta}
\end{align}

\noindent where again we are at $k_{y}=0$ for all the $F$'s in the
above, and the $k$ in the above expresses the $k_{x}$ coordinate.
By the above $F_{-k_{2},-k_{1}}=B(k_{2})F_{k_{1},k_{2}}^{T}B^{\dagger}(k_{1}),$
we have:

\[
F_{-\Delta k,0}=B(\Delta k)F_{0,\Delta k}^{T}B^{\dagger}(0),\]

\begin{equation}
F_{-2\Delta k,-\Delta k}=B(2\Delta k)F_{\Delta k,2\Delta k}^{T}B^{\dagger}(\Delta k),\,\,\cdots
\end{equation}

Hence the Wilson loop above becomes:

\begin{align}
D & =B(\frac{N_{x}}{2}\Delta k)F_{(\frac{N_{x}}{2}-1)\Delta k,\frac{N_{x}}{2}\Delta k}^{T}B^{\dagger}((\frac{N_{x}}{2}-1)\Delta k)\cdots\nonumber \\
 & \times B(2\Delta k)F_{\Delta k,2\Delta k}^{T}B^{\dagger}(\Delta k)B(\Delta k)F_{0,\Delta k}^{T}B^{\dagger}(0)F_{0,\Delta k}\nonumber \\
 & \times F_{\Delta k,2\Delta k}F_{2\Delta k,3\Delta k}\cdots F_{(\frac{N_{x}}{2}-1)\Delta k,\frac{N_{x}}{2}\Delta k}\nonumber \\
 & =B(\frac{N_{x}}{2}\Delta k)F_{(\frac{N_{x}}{2}-1)\Delta k,\frac{N_{x}}{2}\Delta k}^{T}F_{(\frac{N_{x}}{2}-2)\Delta k,(\frac{N_{x}}{2}-1)\Delta k}^{T}\cdots\nonumber \\
 & \times F_{\Delta k,2\Delta k}^{T}F_{0,\Delta k}^{T}B^{\dagger}(0)F_{0,\Delta k}F_{\Delta k,2\Delta k}F_{2\Delta k,3\Delta k}\cdots\nonumber \\
 & \times F_{(\frac{N_{x}}{2}-2)\Delta k,(\frac{N_{x}}{2}-1)\Delta k}F_{(\frac{N_{x}}{2}-1)\Delta k,\frac{N_{x}}{2}\Delta k}
\end{align}

\noindent where we have used the fact that $B^{\dagger}(k)B(k)=I$
(unitary matrix). We hence see that all the intermediate B matrices
vanish with the exception of the ones at the inversion symmetric points
$0$, $\frac{N_{x}}{2}\Delta k$. This is obviously true for any time-reversal
invariant contour. Moreover, it is suggestive that the two left-over
matrices $B^{\dagger}(0)$, $B(\frac{N_{x}}{2}\Delta k)$ be brought
together, so we must commute $B^{\dagger}(0)$ all across the matrix
chain.

The matrix $B(0)$ (and $B(\pi)$) has the property that it is unitary
and antisymmetric, per $B(k)=B^{T}(-k)$ in eq(\ref{eq:bk-bmk}) .
We then have:

\begin{equation}
B(0)=e^{i\theta}\sigma_{2}
\end{equation}

\noindent The matrix $F_{k_{1},k_{2}},$ for $k_{2}-k_{1}\ll\pi,$
as show in eq(\ref{eq:F(i,i+1)}) has the following form:

\begin{equation}
F_{k_{1},k_{2}}^{mn}=\delta_{mn}+A_{k_{1},k_{2}}^{mn}(k_{2}-k_{1})
\end{equation}

\noindent where $k_{2}-k_{1}=\Delta k$. We decompose the $U(2)$
gauge field into its abelian and non-abelian part:

\begin{equation}
A_{k_{1},k_{2}}^{mn}=A_{k_{1},k_{2}}^{U(1)}\delta_{mn}+A_{k_{1},
k_{2}}^{SU(2),i}(\sigma^{i})_{mn}
\end{equation}

\noindent where $i=1,2,3$ and double index implies summation. $A_{k_{1},k_{2}}^{U(1)}$,
$A_{k_{1},k_{2}}^{SU(2),i}$ are numbers. We have:

\begin{align}
 & (F_{k_{1},k_{2}}^{T})^{mn}=(F_{k_{1},k_{2}})^{nm}\nonumber \\
 & =\delta_{mn}+(A_{k_{1},k_{2}}^{U(1)}\delta_{mn}+A_{k_{1},k_{2}}^{SU(2),i}
 (\sigma^{i})_{mn}^{T})(k_{2}-k_{1})
 \end{align}

\noindent We have that:

\begin{align}
(F_{k_{1},k_{2}}^{T})B^{\dagger}(0) & =B^{\dagger}(0)(I+A_{k_{1},k_{2}}^{U(1)}(k_{2}-k_{1})I)\nonumber \\
 & +A_{k_{1},k_{2}}^{SU(2),i}(\sigma^{i})^{T}(k_{2}-k_{1})B^{\dagger}(0)
 \end{align}

\noindent we have commuted the first two terms easily as they are
proportional to the identity, but the last term required a bit more
work:

\[
\sigma_{x}^{T}\sigma_{y}=-\sigma_{y}\sigma_{x}\]

\[
\sigma_{y}^{T}\sigma_{y}=-\sigma_{y}\sigma_{y}\]

\begin{equation}
\sigma_{z}^{T}\sigma_{y}=-\sigma_{y}\sigma_{z}
\end{equation}

and hence:

\begin{align}
 & A_{k_{1},k_{2}}^{SU(2),i}(\sigma^{i})^{T}B^{\dagger}(0)=
 e^{-i\theta}A_{k_{1},k_{2}}^{SU(2),i}(\sigma^{i})^{T}\sigma_{2}\nonumber \\
 & =-e^{-i\theta}\sigma_{2}A_{k_{1},k_{2}}^{SU(2),i}\sigma^{i}=
 -B^{\dagger}(0)A_{k_{1},k_{2}}^{SU(2),i}\sigma^{i}
 \end{align}

Hence:

\begin{align}
 & (F_{k_{1},k_{2}}^{T})B^{\dagger}(0)\nonumber \\
 & =B^{\dagger}(0)[I+(A_{k_{1},k_{2}}^{U(1)}I-A_{k_{1},k_{2}}^{SU(2),i}
 \sigma^{i})(k_{2}-k_{1})]\nonumber \\
 & =B^{\dagger}(0)[I-(A_{k_{1},k_{2}}^{U(1)}I+A_{k_{1},k_{2}}^{SU(2),i}
 \sigma^{i})(k_{2}-k_{1})\nonumber \\
 & +2A_{k_{1},k_{2}}^{U(1)}I(k_{2}-k_{1})]
 \end{align}

We also have

\begin{align}
 & I-(A_{k_{1},k_{2}}^{U(1)}I+A_{k_{1},k_{2}}^{SU(2),i}\sigma^{i})
 (k_{2}-k_{1})+2A_{k_{1},k_{2}}^{U(1)}I(k_{2}-k_{1})\nonumber \\
 & =I+(A_{k_{1},k_{2}})^{\dagger}(k_{2}-k_{1})+2A_{k_{1},k_{2}}^{U(1)}
 (k_{2}-k_{1})I\nonumber \\
 & \approx(I+(A_{k_{1},k_{2}})^{\dagger}(k_{2}-k_{1}))(I+2A_{k_{1},
 k_{2}}^{U(1)}(k_{2}-k_{1}))\nonumber \\
 & \approx F_{k_{1},k_{2}}^{\dagger}e^{2A_{k_{1},k_{2}}^{U(1)}(k_{2}-k_{1})}
 \end{align}

\noindent where in the limit of $k_{2}-k_{1}\ll2\pi$ (the case in
all our terms) we neglect $(k_{2}-k_{1})^{2}$ order terms, and where
$A_{k_{1},k_{2}}$ is the full $U(2)$ field strength. We have also
used the fact that:

\begin{align}
A_{k_{1},k_{2}}^{mn}(k_{2}-k_{1}) & =\langle m,k_{1}|(|n,k_{2}\rangle-|n,k_{1}\rangle)\nonumber \\
 & =\langle m,k_{1}|n,k_{2}\rangle-\delta_{mn}\nonumber \\
 & =\delta_{mn}-\langle m,k_{2}|n,k_{1}\rangle\nonumber \\
 & =-[\langle n,k_{1}|m,k_{2}\rangle-\delta_{mn}]^{*}\nonumber \\
 & =-(A_{k_{1},k_{2}}^{nm})^{*}(k_{2}-k_{1})
 \end{align}

\noindent For $k_{2}-k_{1}\ll2\pi$, to give the above equation, we
used the finite difference version of the equality:

\[
\langle m,k|n,k\rangle=\delta_{mn}\rightarrow(\partial_{k}\langle m,k|)|n,k\rangle+\langle m,k|(\partial_{k}|n,k\rangle)=0\]

which reads by putting:

\[
(\partial_{k_{1}}\langle m,k_{1}|)|n,k_{1}\rangle=\frac{(\langle m,k_{2}|-\langle m,k_{1}|)|n,k_{1}\rangle}{k_{2}-k_{1}}\]
and
\[
\langle m,k_{1}|(\partial_{k_{1}}|n,k_{1}\rangle)=\frac{\langle m,k_{1}|(|n,k_{2}\rangle-|n,k_{1}\rangle)}{k_{2}-k_{1}}\]

\begin{equation}
\langle m,k_{2}|n,k_{1}\rangle+\langle m,k_{1}|n,k_{2}\rangle=2\delta_{mn}
\end{equation}

After all this long detour, we have proved:

\begin{equation}
F_{k_{1},k_{2}}^{T}B^{\dagger}(0)=B^{\dagger}(0)
F_{k_{1},k_{2}}^{\dagger}e^{2A_{k_{1},k_{2}}^{U(1)}(k_{2}-k_{1})}
\end{equation}

We then return to the $U(2)$ Wilson loop.

\begin{align}
D & =B(\frac{N_{x}}{2}\Delta k)F_{(\frac{N_{x}}{2}-1)\Delta k,\frac{N_{x}}{2}\Delta k}^{T}F_{(\frac{N_{x}}{2}-2)\Delta k,(\frac{N_{x}}{2}-1)\Delta k}^{T}\nonumber \\
 & \times F_{(\frac{N_{x}}{2}-3)\Delta k,(\frac{N_{x}}{2}-2)\Delta k}^{T}F_{(\frac{N_{x}}{2}-4)\Delta k,(\frac{N_{x}}{2}-3)\Delta k}^{T}\cdots\nonumber \\
 & \times F_{\Delta k,2\Delta k}^{T}F_{0,\Delta k}^{T}B^{\dagger}(0)F_{0,\Delta k}F_{\Delta k,2\Delta k}F_{2\Delta k,3\Delta k}\cdots\nonumber \\
 & \times F_{(\frac{N_{x}}{2}-2)\Delta k,(\frac{N_{x}}{2}-1)\Delta k}F_{(\frac{N_{x}}{2}-1)\Delta k,\frac{N_{x}}{2}\Delta k}\nonumber \\
 & =B(\frac{N_{x}}{2}\Delta k)F_{(\frac{N_{x}}{2}-1)\Delta k,\frac{N_{x}}{2}\Delta k}^{T}F_{(\frac{N_{x}}{2}-2)\Delta k,(\frac{N_{x}}{2}-1)\Delta k}^{T}\nonumber \\
 & \times F_{(\frac{N_{x}}{2}-3)\Delta k,(\frac{N_{x}}{2}-2)\Delta k}^{T}F_{(\frac{N_{x}}{2}-4)\Delta k,(\frac{N_{x}}{2}-3)\Delta k}^{T}\cdots\nonumber \\
 & \times F_{\Delta k,2\Delta k}^{T}B^{\dagger}(0)e^{2A_{0,\Delta k}^{U(1)}\Delta k}F_{0,\Delta k}^{\dagger}F_{0,\Delta k}F_{\Delta k,2\Delta k}\cdots\nonumber \\
 & \times F_{(\frac{N_{x}}{2}-2)\Delta k,(\frac{N_{x}}{2}-1)\Delta k}F_{(\frac{N_{x}}{2}-1)\Delta k,\frac{N_{x}}{2}\Delta k}\nonumber \\
 & =e^{2(A_{0,\Delta k}^{U(1)}+A_{\Delta k,2\Delta k}^{U(1)}+A_{2\Delta k,3\Delta k}^{U(1)}+\cdots+A_{(N_{x}/2-1)\Delta k,(N_{x}/2)\Delta k}^{U(1)})\Delta k}\nonumber \\
 & \times B(\frac{N_{x}}{2}\Delta k)B^{\dagger}(0)
 \end{align}

\noindent where we have used $U_{0,\Delta k}^{\dagger}U_{0,\Delta k}=U_{\Delta k,2\Delta k}^{\dagger}U_{\Delta k,2\Delta k}=I,\, etc...$.
Hence:

\begin{align}
D & =e^{2(A_{0,\Delta k}^{U(1)}+A_{\Delta k,2\Delta k}^{U(1)}+A_{2\Delta k,3\Delta k}^{U(1)}+\cdots+A_{(N_{x}/2-1)\Delta k,(N_{x}/2)\Delta k}^{U(1)})\Delta k}\nonumber \\
 & \times B(\pi)B^{\dagger}(0)
 \end{align}

We note that the phase above is twice the $U(1)$ (matrix, not traced)
phase picked up from $0$ to $\pi$. Lets re-define it by re-expressing
it from $-\pi$ to $\pi$. i.e. the full abelian Berry for the interval
considered. We have

\begin{equation}
A_{k_{1},k_{2}}^{U(1)}\Delta k=\frac{1}{2}\sum_{n}\langle n,k_{1}|(|n,k_{2}\rangle-|n,k_{1}\rangle)
\end{equation}

\begin{align}
A_{-k_{2},-k_{1}}^{U(1)}\Delta k & =\frac{1}{2}\sum_{n}\langle n,-k_{2}|(|n,-k_{1}\rangle-|n,-k_{2}\rangle)\nonumber \\
 & =\frac{1}{2}\sum_{n}[\langle n,-k_{2}|n,-k_{1}\rangle-\langle n,-k_{2}|n,-k_{2}\rangle]\nonumber \\
 & =\frac{1}{2}tr[F_{-k_{2},-k_{1}}-I]\nonumber \\
 & =\frac{1}{2}tr[B(k_{2})F_{k_{1},k_{2}}^{T}B^{\dagger}(k_{1})-I]
 \end{align}

\noindent where we have used eq(\ref{eq:F(-k)}). Since $k_{2}-k_{1}\ll2\pi$,
we can approximate $B(k_{2})-B(k_{1})$ as small and write $B(k_{2})=B(k_{1})+B(k_{2})-B(k_{1})$
to get:

\begin{align}
 & A_{-k_{2},-k_{1}}^{U(1)}\Delta k\nonumber \\
 & =\frac{1}{2}tr[B(k_{1})F_{k_{1},k_{2}}^{T}B^{\dagger}(k_{1})-I\nonumber \\
 & +(B(k_{2})-B(k_{1}))F_{k_{1},k_{2}}^{T}B^{\dagger}(k_{1})]\nonumber \\
 & =\frac{1}{2}tr[F_{k_{1},k_{2}}^{T}-I+(B(k_{2})-B(k_{1}))
 F_{k_{1},k_{2}}^{T}B^{\dagger}(k_{1})]\nonumber \\
 & =\frac{1}{2}tr[F_{k_{1},k_{2}}-I+(B(k_{2})-B(k_{1}))
 F_{k_{1},k_{2}}^{T}B^{\dagger}(k_{1})]\nonumber \\
 & =A_{k_{1},k_{2}}^{U(1)}\Delta k+\frac{1}{2}tr[(B(k_{2})-B(k_{1}))F_{k_{1},k_{2}}^{T}B^{\dagger}(k_{1})]
 \end{align}

\noindent As $B(k_{2})-B(k_{1})$ is considered small for $k_{2}-k_{1}\ll2\pi$,
we take

\begin{equation}
(B(k_{2})-B(k_{1}))F_{k_{1},k_{2}}^{T}\approx B(k_{2})-B(k_{1})
\end{equation}

\noindent where we took $F_{k_{1},k_{2}}^{T}\approx I$ if multiplied
by another small number. Hence:

\begin{equation}
A_{-k_{2},-k_{1}}^{U(1)}\Delta k=A_{k_{1},k_{2}}^{U(1)}\Delta k+\frac{1}{2}tr[(B(k_{2})-B(k_{1}))B^{\dagger}(k_{1})]
\end{equation}

We then find:

\begin{align}
 & 2(A_{0,\Delta k}^{U(1)}+A_{\Delta k,2\Delta k}^{U(1)}+A_{2\Delta k,3\Delta k}^{U(1)}+\cdots\nonumber \\
 & +A_{(N_{x}/2-1)\Delta k,(N_{x}/2)\Delta k}^{U(1)})\Delta k\nonumber \\
 & =(A_{-(N_{x}/2)\Delta k,-(N_{x}/2-1)\Delta k}^{U(1)}+\cdots\nonumber \\
 & +A_{-2\Delta k,-\Delta k}^{U(1)}+A_{-\Delta k,0}^{U(1)}+A_{0,\Delta k}^{U(1)}+A_{\Delta k,2\Delta k}^{U(1)}\nonumber \\
 & +\cdots+A_{(N_{x}/2-1)\Delta k,(N_{x}/2)\Delta k}^{U(1)})\Delta k\nonumber \\
 & -\frac{1}{2}\intop_{0}^{\pi}dktr[B^{\dagger}(k)\nabla_{k}B(k)]
 \end{align}

\noindent The first term is just the $U(1)$ phase in the contour
direction: $\intop_{-\pi}^{\pi}A^{U(1)}(k)dk$. The Wilson loop is
then:

\begin{equation}
W=e^{\intop_{-\pi}^{\pi}A^{U(1)}(k)dk-\frac{1}{2}\intop_{0}^{\pi}
dktr[B^{\dagger}(k)\nabla_{k}B(k)]}\cdot B(\pi)B^{\dagger}(0)
\end{equation}

\noindent As $B(k)$ is unitary, we also know that:

\begin{equation}
tr[B^{\dagger}(k)\nabla_{k}B(k)]=\nabla_{k}log\, det\, B(k)
\end{equation}

and hence:

\begin{equation}
e^{-\frac{1}{2}\intop_{0}^{\pi}dktr[B^{\dagger}(k)\nabla_{k}B(k)]}
=e^{-\frac{1}{2}log[\frac{detB(\pi)}{detB(0)}]}=\sqrt{\frac{detB(0)}{detB(\pi)}}
\end{equation}

The Wilson loop becomes:

\begin{equation}
D=e^{\intop_{-\pi}^{\pi}A^{U(1)}(k)dk}\cdot\sqrt{\frac{detB(0)}{detB(\pi)}}\cdot B(\pi)B^{\dagger}(0)
\end{equation}

As we said before, $B(0)$, $B(\pi)$ are unitary, $2$ by 2 matrices,
antisymmetric, so:

\begin{equation}
B(0)=Pf(B(0))\left[\begin{array}{cc}
0 & 1\\
-1 & 0\end{array}\right]
\end{equation}

\begin{equation}
B(\pi)=Pf(B(\pi))\left[\begin{array}{cc}
0 & 1\\
-1 & 0\end{array}\right]
\end{equation}

\noindent where Pf is the pfaffian of the matrix. We hence have:

\begin{equation}
D=e^{\intop_{-\pi}^{\pi}A^{U(1)}(k)dk}\sqrt{\frac{detB(0)}
{detB(\pi)}}\frac{Pf(B(\pi))}{Pf(B(0))}I
\end{equation}
 where $I$ is the $2\times2$ identity matrix.

We now make several observations. Obviously, the above equality is
valid on both time-reversal invariant lines at $k_{y}=0$ and $k_{y}=\pi$,
i.e. we can define two Wilson loops:

\begin{eqnarray}
 & D(k_{y}=0)=e^{\intop_{-\pi}^{\pi}A^{U(1)}(k_{x},k_{y}=0)dk_{x}}\nonumber \\
 & \times\sqrt{\frac{detB(0,0)}{detB(\pi,0)}}\frac{Pf(B(\pi,0))}{Pf(B(0,0))}I
 \end{eqnarray}
 and

\noindent \begin{eqnarray}
 & D(k_{y}=\pi)=e^{\intop_{-\pi}^{\pi}A^{U(1)}(k_{x},k_{y}=\pi)dk_{x}}\nonumber \\
 & \times\sqrt{\frac{detB(0,\pi)}{detB(\pi,\pi)}}\frac{Pf(B(\pi,\pi))}{Pf(B(0,\pi))}I
 \end{eqnarray}
 Second, we notice that the $U(1)$ phase factor is not just the usual
abelian Berry phase but only \emph{half} of it. Indeed, as per our
definition: 
\begin{equation}
A_{k_{1},k_{2}}^{m,n}=A_{k_{1},k_{2}}^{U(1),i}\delta_{mn}
+A_{k_{1},k_{2}}^{SU(2),i}(\sigma^{i})_{mn}
\end{equation}
 This implies:

\noindent \begin{equation}
A_{\vec{k}}^{U(1),i}=\frac{1}{2}\sum_{m}\langle m,\vec{k}|m,\vec{k}\rangle
\end{equation}
 which has a $1/2$ difference from the usual form. This difference
is actually important. Define: \begin{equation}
\Phi(k_{y})=\oint_{-\pi}^{\pi}dk_{x}A_{x}(k_{x},k_{y})=\log\det D(k_{y})
\end{equation}
 We then have: \begin{align}
\int_{0}^{\pi}\nabla_{k_{y}}\Phi(k_{y}) & =\int_{0}^{\pi}\nabla_{k_{y}}\log\det D(k_{y})\nonumber \\
 & =\Phi(\pi)-\Phi(0)+2\pi iM_{n}
 \end{align}
 where $M_{n}$ is the winding number of the phase $\Phi(\pi)$. The
phase $\Phi(k_{y})$ is the sum of the phases $\phi_{1}(k_{y})$ and
$\phi_{2}(k_{y})$ of the two eigenvalues of the Wilson loop \emph{both
defined in the interval} $[0,2\pi]$. Each of these eigenvalues has
a winding number which adds to $M_{n}$, and the system will turn
to be nontrivial if the system has a odd $M_{n}$. We now take the
Wilson loop $W$ from $k_{x}=-\pi,\pi$ at $k_{y}=0$, and then from
$k_{x}=\pi,-\pi$ at $k_{y}=\pi$:

\noindent \begin{eqnarray}
W & = & D(k_{y}=0)(D(k_{y}=\pi))^{-1}\nonumber \\
 & = & e^{\frac{1}{2}(\Phi(0)-\Phi(k))}\times\sqrt{\frac{detB(0,0)}{detB(\pi,0)}}
 \frac{Pf(B(\pi,0))}{Pf(B(0,0))}\nonumber \\
 & \times & \sqrt{\frac{detB(0,\pi)}{detB(\pi,\pi)}}\frac{Pf(B(\pi,\pi))}{Pf(B(0,\pi))}I\nonumber \\
 & = & e^{\frac{1}{2}(-\int_{0}^{\pi}\nabla_{k_{y}}\log\det D(k_{y}))+\pi iM_{n}}\nonumber \\
 & \times & \sqrt{\frac{detB(0,0)}{detB(\pi,0)}}\frac{Pf(B(\pi,0))}{Pf(B(0,0))}\nonumber \\
 & \times & \sqrt{\frac{detB(0,\pi)}{detB(\pi,\pi)}}\frac{Pf(B(\pi,\pi))}{Pf(B(0,\pi))}I
 \end{eqnarray}
 As such:

\begin{align}
D(k_{y} & =0)e^{\frac{1}{2}(\int_{0}^{\pi}\nabla_{k_{y}}\log\det D(k_{y}))}(D(k_{y}=\pi))^{-1}\nonumber \\
= & e^{\pi iM_{n}}\sqrt{\frac{detB(0,0)}{detB(\pi,0)}}\frac{Pf(B(\pi,0))}{Pf(B(0,0))}\nonumber \\
\times & \sqrt{\frac{detB(0,\pi)}{detB(\pi,\pi)}}\frac{Pf(B(\pi,\pi))}{Pf(B(0,\pi))}I
\end{align}

\noindent We have proved that both $D(k_{y}=0)$ and $D(k_{y}=\pi)$
are proportional to unity matrix, up to a sign. For a smooth gauge,
the difference in sign is taken by the contour term $e^{\frac{1}{2}(\int_{0}^{\pi}\nabla_{k_{y}}\log\det D(k_{y}))}$
to give

\noindent \begin{equation}
D(k_{y}=0)e^{\frac{1}{2}(\int_{0}^{\pi}\nabla_{k_{y}}\log\det D(k_{y}))}(D(k_{y}=\pi))^{-1}=I\end{equation}
 to give 
 \begin{equation}
1=e^{\pi iM_{n}}\sqrt{\frac{detB(0,0)}{detB(\pi,0)}}\frac{Pf(B(\pi,0))}
{Pf(B(0,0))}\sqrt{\frac{detB(0,\pi)}{detB(\pi,\pi)}}\frac{Pf(B(\pi,\pi))}
{Pf(B(0,\pi))}
\end{equation}

\noindent which says that the pfaffian invariant is just the parity
of the band switch number $M_{n}$. For $M_{n}$ odd, it is nontrivial.
Note that although our proof above is explicit only for two occupied
bands, it can be easily extended to the $2N_{occupied}$ band case
when we realize that that case is just a tensor product (upon removing
accidental degeneracies) of $N_{occupied}$ time-reversal invariant
multiplets for which the above expression applies.

\section{Wilson loop and the $Z_{2}$ invariant expressed as an obstruction}

An alternative formulation of the $Z_{2}$ invariant has been defined
by Fu and Kane\cite{fu_time_2006}, where the $Z_{2}$ invariant is
expressed as an obstruction of the $U(1)$ Berry's phase gauge field
in half of the Brillouin zone. This approach has some convenience
in its similarity with the Chern number formula of the quantum Hall
states by Thouless \textit{et al}\cite{thouless_quantized_1982}.
The application of this approach to numerical calculation of the $Z_{2}$
invariant in finite size systems has been studied by Fukui and Hatsugai\cite{fukui_quantum_2007}
. Here we provide an alternative proof of the relation between our
Wilson loop approach and the $Z_{2}$ invariant through the obstruction
formula. We start by reviewing the obstruction formulation of Fu and
Kane\cite{fu_time_2006} (Appendix A1). Consider $|n,k\rangle$ as
the occupied Bloch bands. We make the gauge choice \begin{eqnarray}
|n,-k\rangle=\mathcal{T}_{nm}T(|m,k\rangle)\label{gaugechoice}\end{eqnarray}
 with $\mathcal{T}$ an antisymmetric matrix satisfying $\mathcal{T}^{2}=-1$.
{Comparing to Eq. (\ref{eqn:Ttrans}), the gauge choice here corresponds
to the requirement that $B_{nm}(k)$ is independent from $k$.} More
explicitly, with $2N$ occupied bands we can label the bands in pairs
as $|n,k\rangle$ and $|\bar{n},k\rangle$ with $n=1,...,N$. Time-reversal
acts as $|\bar{n},-k\rangle=T|n,k\rangle$, $|n,-k\rangle=-T|\bar{n},k\rangle$.
so that the wavefunctions in the lower half Brillouin zone (BZ) defined
by $k_{y}\in[-\pi,0]$ is determined by those in the upper half BZ,
denoted by $\tau_{1/2}$. With this gauge choice, for topological
insulator it is not possible to make a continuous and single-valued
choice of the wavefunctions in the whole Brillouin zone. However,
it is always possible to define the wavefunctions continuously in
the half BZ $\tau_{1/2}$, so that all obstructions are pushed to
the boundary between the two half BZs, \textit{i.e.} the two lines
$k_{y}=0$ and $k_{y}=\pi$. In such a gauge choice, Fu and Kane shows
that the $Z_{2}$ invariant is given by an obstruction in the half
Brillouin zone: \begin{eqnarray}
\Delta=\frac{1}{2\pi}\left(\oint_{\partial\tau_{1/2}}{\bf A\cdot dl}-\int_{\tau_{1/2}}d^{2}kF_{xy}\right)~{\rm ~mod~}2\end{eqnarray}
 where $A_{i}=-i\sum_{n}\langle nk|\partial_{i}|nk\rangle$ is the
$U(1)$ part of the Berry phase connection. By first integrating over
$k_{x}$ and define the $U(1)$ Wilson loop \begin{eqnarray}
\Phi(k_{y})=\oint_{-\pi}^{\pi}dk_{x}A_{x}(k_{x},k_{y})\end{eqnarray}
 The $Z_{2}$ invariant can be expressed as

\begin{align}
\Delta & =\frac{1}{2\pi}\left(\int_{0}^{\pi}dk_{y}\partial_{k_{y}}
\Phi(k_{y})-\left(\Phi(\pi)-\Phi(0)\right)\right){\rm ~mod~}2\label{eq:z2}
\end{align}

\begin{figure}[b]
 \includegraphics[scale=0.35]{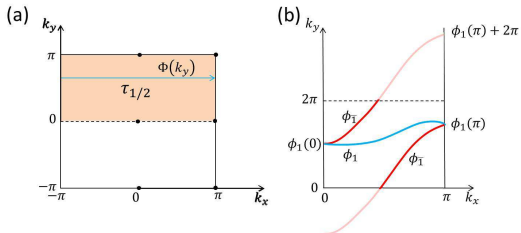} \caption{(a) Definition of the half Brillouin zone and the paths to define
Wilson loops. (b) Schematic picture of eigenvalues $\phi_{n}(k_{y})$
for a two band model. $\phi_{1}(k_{y})$ has winding number $0$ and
$\phi_{\bar{1}}(k_{y})$ has winding number $1$, when $\phi_{1}(0)$
and $\phi_{1}(\pi)$ are chosen to be in $[0,2\pi)$. }

\label{fig1}
\end{figure}

If we don't have any restriction on the gauge choice (other than requiring
that the wavefunctions are continuous in $\tau_{1/2}$ so that $\Phi(k_{y})$
is continuous and well-defined), $\Delta$ can be $0$, or any arbitrary
integer. However, the gauge choice eq.(\ref{gaugechoice}) removes
this ambiguity. Consider the states $|nk\rangle$ for $k_{y}=0$.
A gauge transformation \begin{eqnarray}
|nk\rangle\rightarrow e^{i\varphi_{{\bf k}}}|nk\rangle\end{eqnarray}
 corresponds to a gauge transformation ${\bf a}\rightarrow{\bf a}+\nabla_{{\bf k}}\varphi$,
which leads to the change in the flux 
\begin{eqnarray}
\Phi(k_{y}=0)\rightarrow\Phi(k_{y}=0)+2N\oint_{-\pi}^{\pi}dk_{x}\partial_{x}\varphi_{k_{x},0}
\end{eqnarray}
 However, to preserve the condition eq.(\ref{gaugechoice}) we have
to require \begin{eqnarray}
\varphi_{{\bf k}}=-\varphi_{-{\bf k}}\label{gaugetrans}\end{eqnarray}
 so that 
 \begin{eqnarray}
\oint_{-\pi}^{\pi}dk_{x}\partial_{x}\varphi_{k_{x},0}=2\int_{0}^{\pi}
\partial_{x}\varphi_{k_{x},0}=2\left(\varphi_{\pi,0}-\varphi_{0,0}\right)
\end{eqnarray}
 Since we also have $\varphi_{\pi,0}$ and $\varphi_{0,0}=0{\rm ~mod~}\pi$
from eq.(\ref{gaugetrans}), the allowed change of $\Phi(k_{y}=0)$
in the gauge transformations that preserves the gauge choice eq.(\ref{gaugechoice})
can only be 

\begin{eqnarray}
\Phi(k_{y}=0) & \rightarrow & \Phi(k_{y}=0)+4N\left(\varphi_{\pi,0}-\varphi_{0,0}\right)\nonumber \\
 & = & \Phi(k_{y}=0)+4\pi n,~n\in\mathbb{Z}
 \end{eqnarray}
 The same is true for $\Phi(k_{y}=\pi)$. Consequently, $\Phi(k_{y}=0)$
and $\Phi(k_{y}=\pi)$ are well-defined modular $4\pi$, so that the
$Z_{2}$ quantity $\Delta$ defined by eq.(\ref{eq:z2}) is well-defined.

Now we relate this result to the non-Abelian Wilson loop. When there
are $2N$ bands occupied, a $U(2N)$ Berry phase gauge field $a_{i}^{nm}=-i\langle nk|\partial_{i}|mk\rangle$
is defined. We can define the $U(2N)$ Wilson loop along the same
equal-$k_{y}$ loops: 

\begin{eqnarray}
W(k_{y})=Pe^{i\oint dk_{x}a_{x}(k_{x},k_{y})}\in U(2N)\end{eqnarray}
 The $U(1)$ gauge field is related to the $U(2N)$ gauge field by
$A_{i}={\rm Tr}a_{i}$ so that the $U(1)$ flux $\Phi(k_{y})$ is
related to $W(k_{y})$ as $e^{i\Phi(k_{y})}=\det W(k_{y})$. Denote
the eigenvalues of $W(k_{y})$ as $e^{i\phi_{n}(k_{y})}$ with $n=1,..,2N$,
we have $\Phi(k_{y})=\sum_{n}\phi_{n}(k_{y})~{\rm ~mod~}2\pi$. Thus

\begin{eqnarray}
\Delta & = & \frac{1}{2\pi}\left(\sum_{n}\int_{0}^{\pi}dk_{y}\partial_{k_{y}}
\phi_{n}(k_{y})-\sum_{n}\left(\phi_{n}(\pi)-\phi_{n}(0)\right)\right)\nonumber \\
 &  & {\rm ~mod~}2\label{eq:z2forU2N}
 \end{eqnarray}
 Now we study the effect of the gauge choice eq.(\ref{gaugechoice})
on the $U(2N)$ gauge field.

\begin{align}
{\bf a}_{-{\bf k}}^{nm} & =-i\langle n,-k|\nabla_{-{\bf k}}|m,-k\rangle\nonumber \\
 & =i\mathcal{T}_{ml}\mathcal{T}_{np}T\left(\langle pk|\nabla_{{\bf k}}|lk\rangle\right)\nonumber \\
 & =\mathcal{T}_{ml}\mathcal{T}_{np}\left({\bf a}_{{\bf k}}^{pl}\right)^{*}=\left(\mathcal{T}{\bf a}_{{\bf k}}^{T}\mathcal{T}^{-1}\right)_{nm}\label{eq:Tsym}\\
\Rightarrow W(-k_{y}) & =\mathcal{T}W^{T}(k_{y})\mathcal{T}^{-1}
\end{align}

Thus for $k_{y}=0$ or $\pi$, $e^{i\phi_{n}(k_{y})}$ is doubly degenerate.
If we label a pair of degenerate eigenvalues by $n$ and $\bar{n}$,
the gauge choice eq.(\ref{gaugechoice}) corresponds to the choice
of $\phi_{n}(k_{y})=\phi_{\bar{n}}(k_{y})$. Indeed we see that if
we make this choice, an ambiguity of $2\pi$ in $\phi_{n}$ leads
to an ambiguity of $4\pi$ in $\sum_{n}\phi_{n}$. Thus $\Delta$
defined in eq.(\ref{eq:z2forU2N}) is well-defined. To simplify the
formula, we choose $\phi_{n}(0)$ and $\phi_{n}(\pi)$ to be in $[0,2\pi)$.
Thus

\begin{eqnarray}
\int_{0}^{\pi}dk_{y}\partial_{k_{y}}\phi_{n}(k_{y})=\phi_{n}(\pi)-\phi_{n}(0)+2\pi M_{n}
\end{eqnarray}

 with $M_{n}$ the winding number of phase $\phi_{n}$, which is equal
to the number of times $\phi_{n}$ crosses the line $\phi_{n}=2\pi$
from below. For example in Fig.\ref{fig1} (b) $\phi_{1}$ has winding
number $0$ and $\phi_{\bar{1}}$ has winding number $1$. In this
way we get 

\begin{eqnarray}
\Delta=\sum_{n}M_{n}~{\rm ~mod~}2
\end{eqnarray}

 The number $\sum_{n}M_{n}$ simply counts how many eigenvalues $\phi_{n}$
crosses $\phi=2\pi$ line (or any other reference line) from below.
Thus the $Z_{2}$ invariant is simply determined by the parity of
the number of eigenvalue curves $\phi_{n}(k_{y})$ which crosses a
reference line $\phi={\rm constant}$.

\bibliographystyle{apsrev}
\bibliography{TIs}

\end{document}